\pdfoutput=1
\documentclass{emulateapj}
\usepackage{graphics,graphicx}
\usepackage{fancyheadings}
\usepackage{color}
\usepackage{natbib}
\usepackage{times}
\usepackage{hyperref}
\usepackage{breakurl} 

\newcommand{\as}{$^{\prime\prime}$}
\newcommand{\am}{$^{\prime}$}
\newcommand{\Lu}{\,erg\,s$^{-1}$}
\newcommand{\Su}{\,erg\,cm$^{-2}$\,s$^{-1}$}

\newcommand{\Uu}{\,erg\,cm$^{-3}$}
\newcommand{\Ms}{\mathcal{M}_{sh}}
\newcommand{\rg}{PKS\,B1358$-$113}
\newcommand{\cl}{Abell\,1836}
\newcommand{\cxo}{{\it Chandra}}
\newcommand{\xmm}{XMM-{\it Newton}}
\newcommand{\ros}{{\it ROSAT}}
\newcommand{\vla}{{\it VLA}}
\newcommand{\hst}{{\it HST}}
\newcommand{\wht}{{\it WHT}}

\def\deg{\hbox{$^\circ$}}

\shorttitle{PKS\,B1358$-$113 in Galaxy Cluster Abell\,1836}
\shortauthors{Stawarz et al.}

\begin{document}

\title{On the Interaction of the PKS\,B1358$-$113 Radio Galaxy\\ with the Abell\,1836 Cluster}

\author{{\L}.~Stawarz$^{1,\,2}$, A.~Szostek$^{3,\,2}$, C.C.~Cheung$^4$, A.~Siemiginowska$^5$, D.~Kozie\l -Wierzbowska$^{2}$, N.~Werner$^3$, A.~Simionescu$^1$, G~.Madejski$^6$, M.C.~Begelman$^7$, D.E.~Harris$^5$, M.~Ostrowski$^2$,\\ and K. Hagino$^1$}

\medskip

\affil{$^1$ Institute of Space and Astronautical Science JAXA, 3-1-1 Yoshinodai, Chuo-ku, Sagamihara, Kanagawa 252-5210, Japan} 
\affil{$^2$ Astronomical Observatory, Jagiellonian University, ul. Orla 171, 30-244 Krak\'ow, Poland}
\affil{$^3$ KIPAC, Stanford University, 452 Lomita Mall, Stanford, CA 94305, USA, and Department of Physics, Stanford University, 382 Via Pueblo Mall, Stanford, CA 94305-4060, USA}
\affil{$^4$ Space Science Division, Naval Research Laboratory, Washington, DC 20375, USA}
\affil{$^5$ Harvard Smithsonian Center for Astrophysics, 60 Garden St, Cambridge, MA 02138, USA}
\affil{$^6$ W. W. Hansen Experimental Physics Laboratory, Kavli Institute for Particle Astrophysics and Cosmology, Department of Physics and SLAC National Accelerator Laboratory, Stanford University, Stanford, CA 94305, USA}
\affil{$^7$ JILA, University of Colorado and National Institute of Standards and Technology, 440 UCB, Boulder, CO 80309-0440, USA}

\medskip

\email{email: {\tt stawarz@astro.isas.jaxa.jp}}
\label{firstpage}

\begin{abstract}
Here we present the analysis of multifrequency data gathered for the Fanaroff-Riley type-II (FR\,II) radio galaxy \rg, hosted in the brightest cluster galaxy in the center of \cl. The galaxy harbors one of the most massive black holes known to date and our analysis of the acquired optical data reveals that this black hole is only weakly active, with a mass accretion rate $\dot{M}_{\rm acc} \sim 2 \times 10^{-4} \, \dot{M}_{\rm Edd} \sim 0.02 \, M_{\odot}$\,yr$^{-1}$.  Based on analysis of new \cxo\ and \xmm\ X-ray observations and archival radio data, and assuming the well-established model for the evolution of FR II radio galaxies, we derive the preferred range for the jet kinetic luminosity $L_{\rm j} \sim (1-6) \times 10^{-3}\, L_{\rm Edd} \sim (0.5-3) \times 10^{45}$\,\Lu. This is above the values implied by various scaling relations proposed for radio sources in galaxy clusters, being instead very close to the maximum jet power allowed for the given accretion rate. We also constrain the radio source lifetime as $\tau_{\rm j} \sim 40-70$\,Myrs, meaning the total amount of deposited jet energy $E_{\rm tot} \sim (2-8) \times 10^{60}$\,ergs. We argue that approximately half of this energy goes into shock-heating of the surrounding thermal gas, and the remaining $50\%$ is deposited into the internal energy of the jet cavity. The detailed analysis of the X-ray data provides indication for the presence of a bow-shock driven by the expanding radio lobes into the \cl\ cluster environment. We derive the corresponding shock Mach number in the range $\mathcal{M}_{sh} \sim 2-4$, which is one of the highest claimed for clusters or groups of galaxies. This, together with the recently growing evidence that powerful FR\,II radio galaxies may not be uncommon in the centers of clusters at higher redshifts, supports the idea that jet-induced shock heating may indeed play an important role in shaping the properties of clusters, galaxy groups, and galaxies in formation. In this context, we speculate on a possible bias against detecting stronger jet-driven shocks in poorer environments, resulting from an inefficient electron heating at the shock front, combined with a relatively long electron-ion temperature equilibration timescale.
\end{abstract}


\section{Introduction}
\label{intro}

One of the most widely debated problems in modern astrophysics is related to heating of the hot gaseous environment in galaxy clusters, which is required by the observed temperature and entropy profiles of the intracluster medium (ICM). This problem is manifested by the apparent lack of observed low-temperature gas ($< T_{\rm vir}/3$, where $T_{\rm vir}$ is the virial temperature) falling at very high rates ($\sim 100-1000\,M_{\odot}$\,yr$^{-1}$) onto the centers of the so-called ``cooling-flow'' systems \citep[see][and references therein]{fab94,pet06}. Although the debate regarding the exact physical processes at work is in this context still ongoing, mechanical heating by relativistic jets and lobes expanding from active galactic nuclei (AGN) of giant ellipticals located in cluster centers (hereafter `brightest cluster galaxies', BCGs), is considered as one of the most promising scenarios \citep[see, e.g., the topical reviews by][and references therein]{mcn07,mcn12}. This idea is supported by the finding that BCGs are indeed typically radio-loud \citep[e.g.,][]{bur90,bes05,crof07}, and that the jets and lobes produced in those systems are, on average, capable of suppressing the dramatic ICM cooling as far as the total available energy release is considered \citep[e.g.,][but see the discussion below]{bir04,raf06,dun08}.

The impact of large-scale jets and lobes on their environment is not restricted to quenching of cooling flows in clusters. Radio-loud AGN are now understood to play a crucial role also in altering the gas properties in galaxy groups (where, in fact, most of the galaxies in the Universe reside) and isolated massive ellipticals \citep[see][for a review]{mat03,sun12}, or even to suppress star-formation processes in the interstellar medium (ISM) of systems in formation (particularly at high redshifts), influencing in this way the co-evolution of galaxies and central supermassive black holes (SMBHs) in a complex feedback loop \citep{bow06,cat06,crot06}. Hence, understanding the energetics of AGN jets and lobes, as well as how exactly they interact with the ambient medium, is crucial for understanding the structure formation in the Universe in general.

The first models of the ICM heating by AGN jets involved energy dissipation at strong shocks driven in the ambient medium by the expanding jet cocoons, i.e., lobes \citep{cla97,hei98,kai99}. Such shocks were naturally expected in the framework of widely accepted evolutionary scenarios developed for `classical doubles' \citep{sch74,beg89,kai97}. However, while the cavities in the X-ray emitting cluster gas at the positions of radio lobes of BCG-hosted AGN were being discovered already in \ros\ data (see \citealt{boe93} for Perseus\,A, \citealt{car94} for Cygnus\,A, or \citealt{hua98} for Abell\,4059), indicating clearly that the expanding jets do displace the ICM out of the cluster central regions, no radiative signatures for the expected shocks heating up the X-ray emitting gas around the boundaries of the radio lobes were found even in the first high-resolution \cxo\ observations of systems like Hydra\,A \citep{mcn00}, Perseus\,A \citep{fab00}, Abell\,2052 \citep{bla01}, RBS\,797 \citep{sch01}, or Abell\,4059 \citep{hei02}. 

This absence of evidence for strong shocks at the lobes' edges led to the conclusion that the expected supersonic expansion of jet cocoons takes place only in the earliest phases of the source evolution (basically only when the radio structure is still confined within the ISM of the host), and is quickly replaced by a transonic expansion as soon as the lobes reach the ICM scale \citep{rey01}. This transonic expansion of the lobes, combined with the anticipated duration of the jet lifetime/jet duty cycle of the order of $\sim 10-100$\,Myr, which is much shorter than the cluster lifetime, led next to the idea that, after switching off the jet activity, the lobes detach from the central AGN and rise buoyantly in the stratified cluster atmosphere in the form of low-density `bubbles.' The rising bubbles uplift the cool gas from the center, converting their enthalpy into the gas kinetic energy, which may next be thermalized \emph{somehow} in the bubbles' wake \citep{chu01,chu02}. This scenario seemed supported by the discovery of `ghost cavities' \citep[e.g.,][]{fab00}, and also by numerical simulations of the jet evolution in rich cluster environment (e.g., \citealt{bru02,rey02,rus04}; although one should note that the stability of such structures against various types of magneto-hydrodynamic instabilities, which depends crucially on the hardly known ICM magnetic field configuration, is an open issue, e.g., \citealt{jon05,die08,one09}).

Later, however, the evidence for the presence of \emph{weak} shocks driven by expanding lobes in the gaseous atmospheres of clusters, galaxy groups, and isolated systems, started to emerge in deep and very deep \cxo\ and \xmm\ exposures (see \S\,\ref{sh} below and references therein), and placed increased attention back again on the possibility of shock heating of the ICM \citep{voi05}. Yet highly over-pressured jet cocoons expanding supersonically are expected in luminous classical doubles, i.e. Fanaroff-Riley type II sources (FR\,IIs), rather than in low-power Fanaroff-Riley type I radio galaxies (FR\,Is) typically found in the centers of rich clusters \citep[see in this context][]{bru07,guo10,per11,har13}. In fact, a number of observational studies indicated that FR\,IIs at low redshifts avoid dense cluster environment, and that there is a clear difference in richnesses of clusters harboring FR\,I and FR\,IIs central radio galaxies \emph{on average}, albeit with a substantial dispersion \citep[e.g.,][]{lon79,pre88,all93,wan96,zir97,har02,sle08,win11}. There are however strong indications that this may not hold in the higher-$z$ Universe, where luminous FR\,IIs are found also in richer systems, equivalent to Abell class I or higher \citep{yat89,hil91,sie05,bel07,ant12}.

According to \citet{fan74}, radio galaxies with total spectral luminosity density exceeding $P_{\rm 1.4\,GHz} \sim 10^{25}$\,W\,Hz$^{-1}$ are characterized almost exclusively by a `classical-double', edge-brightened morphology, consisting of collimated jets terminating in well-localized bright hotspots, and surrounded by prominent lobes. On the other hand, the overwhelming majority of sources with 1.4\,GHz spectral power below this critical value possess edge-darkened large-scale radio structures with less-collimated jets lacking any clear termination points, but instead extend in a plume-like fashion to larger distances, or form amorphous bubbles of radio-emitting plasma. These two types of objects are referred to as the aforementioned FR\,II and FR\,I type radio galaxies, respectively. \citet{led96} later argued that the FR\,II/FR\,I division depends not only on the total radio power of a source, but also on the properties of the galactic hosts, and that the borderline radio luminosity in particular scales with the host optical luminosity as $\propto L_{\rm opt}^{1.8}$. 

Later investigations showed however that the \citeauthor{led96} scaling is not an absolute separation, as there is a substantial scatter of FR\,IIs around the proposed borderline \citep[e.g.,][]{lin10,koz11,win11}. Still, even only a rough positive dependance of the jet radiative luminosity on the host luminosity is a strong indication that the difference in large-scale radio morphology of a source is due to a combination of the jet kinetic power (scaling somehow with the lobes' radio luminosity) and the properties of the ambient medium (on the galactic scale), rather than due to the jet power alone \citep[see in this context also][]{gop00,kai07,kaw09}. Namely, for a given pressure and density of the ambient medium, the jet has to be powerful enough to pierce through the surrounding gas and to form a strong termination shock (observed as a hotspot) at the tip of an outflow, through which the jet plasma passes by, back-flowing into an over-pressured cocoon/lobe surrounding the jet; the lobe expands next sideways with supersonic velocity driving weaker bow-shock all along its edges.

We note that the FR\,II and FR\,I populations seem to differ also in the AGN emission line properties whereby sources with prominent emission lines (`high-excitation radio galaxies'; HERGs) are typically associated with FR\,II-type radio structures, while majority of FR\,I-type sources possess nuclei with no strong emission lines \citep[`low-excitation radio galaxies'; LERGs; see][]{lai94}. Again, the division is not a strict one, since many FR\,IIs are classified as LERGs, and some FR\,Is are found to have HERG-like nuclei \citep[e.g.,][]{har07,but10,bes12,gen13,ine13}. Despite this caveat, the established correspondence is in general consistent with the idea that the production of powerful jets --- like the ones found in FR\,IIs --- requires higher accretion rates (evidenced by prominent AGN emission lines), with the commonly anticipated borderline at the accretion luminosities in Eddington units, $L_{\rm acc}/L_{\rm Edd} > 0.01$ \citep[see][]{ghi01}. This issue is then particularly interesting in the context of BCG-hosted jets, which are often assumed to be powered only by a hot rarefied gas accreting at a limited rate \citep{all06,mer07,mcn11,rus13}, even though large amounts of cold gas capable of an efficient fueling of central SMBHs are now being routinely found in many giant ellipticals at the centers of evolved clusters \citep[see, e.g.,][]{don11}.

Studying classical doubles in rich cluster environment is therefore important for several reasons, and in particular for understanding (i) the AGN activity and jet duty cycle in the evolved systems, (ii) the evolution of relativistic outflows and their interactions with the ambient medium, as well as (iii) the mechanical heating of the ICM by the expanding AGN jets and lobes. Thus motivated, here we present the analysis of multifrequency data for the FR\,II source \rg\ located in the center of the \cl\ cluster. The paper is organized as follows. In \S\,\ref{data} we describe the target selection, and present the analysis of the arc-second resolution radio maps from the Very Large Array archive, along with the newly acquired optical spectra from the William Herschel Telescope, as well as \cxo\ and \xmm\ X-ray data. In \S\,\ref{results} we discuss the main results of the data analysis for the \rg/\cl\ system. In \S\,\ref{discuss} we summarize our findings in a broader context of the `radio-mode' feedback operating in galaxy clusters. We assumed a standard cosmology with $H_0=71$\,km\,s$^{-1}$\,Mpc$^{-1}$, $\Omega_m=0.27$, and $\Omega_{\Lambda}=0.73$, so that the redshift of the target $z = 0.0363$ corresponds to the luminosity distance of 158\,Mpc and the conversion scale of $0.71$\,kpc/\as.

\section{Target Selection and Data Analysis}
\label{data}

\subsection{Target Selection}
\label{select}

As mentioned in \S\,\ref{intro} above, powerful radio galaxies of the FR\,II type at low redshifts are typically found in relatively poor groups of galaxies, or even in isolated fields, and those associated with richer environments are located rather at the outskirts of merging systems (like for example 3C\,353 in the Zw\,1718.1$-$0108 cluster studied in detail with \cxo\ and \xmm\ by \citealt{kat08} and \citealt{goo08}, respectively). There are only a few well-studied exceptions from this rule, including the famous Cygnus\,A radio galaxy (see \S\,\ref{sh} below and references therein). To search for more `Cygnus\,A-type' cluster radio sources, we examined the results of the NRAO\footnote{The National Radio Astronomy Observatory is a facility of the National Science Foundation operated under cooperative agreement by Associated Universities, Inc.} Very Large Array (\vla) imaging survey of Abell clusters carried out by F.N.~Owen and collaborators \citep{owe92,owe97}. We identified 12 radio galaxies displaying FR\,II type morphology in their sample of $\simeq 400$ clusters at $z < 0.25$. Most of these display large offsets from the cluster centers (3\am--8\am), as expected, and only two are clearly hosted by the brightest cluster galaxies: 4C+67.13 in Abell\,578 ($z = 0.0866$, richness $R = 0$), and \rg\ in \cl\ ($z = 0.0363$, $R=0$)\footnote{Bautz-Morgan cluster morphology classification II; the Abell richness count 41 \citep{lai03}.}. These objects are thus the two clearest, if not only, examples of FR\,II radio galaxies known at the centers of local ($z < 0.25$) Abell clusters with richness $R \geq 0$ (at the northern declinations visible to the \vla). In this paper we present the results of the multifrequency data analysis for the latter target; the 4C+67.13/Abell\,578 system will be discussed in a forthcoming paper (Hagino et al., in prep). 

\begin{figure}[!t]
\begin{center}
\includegraphics[width=\columnwidth]{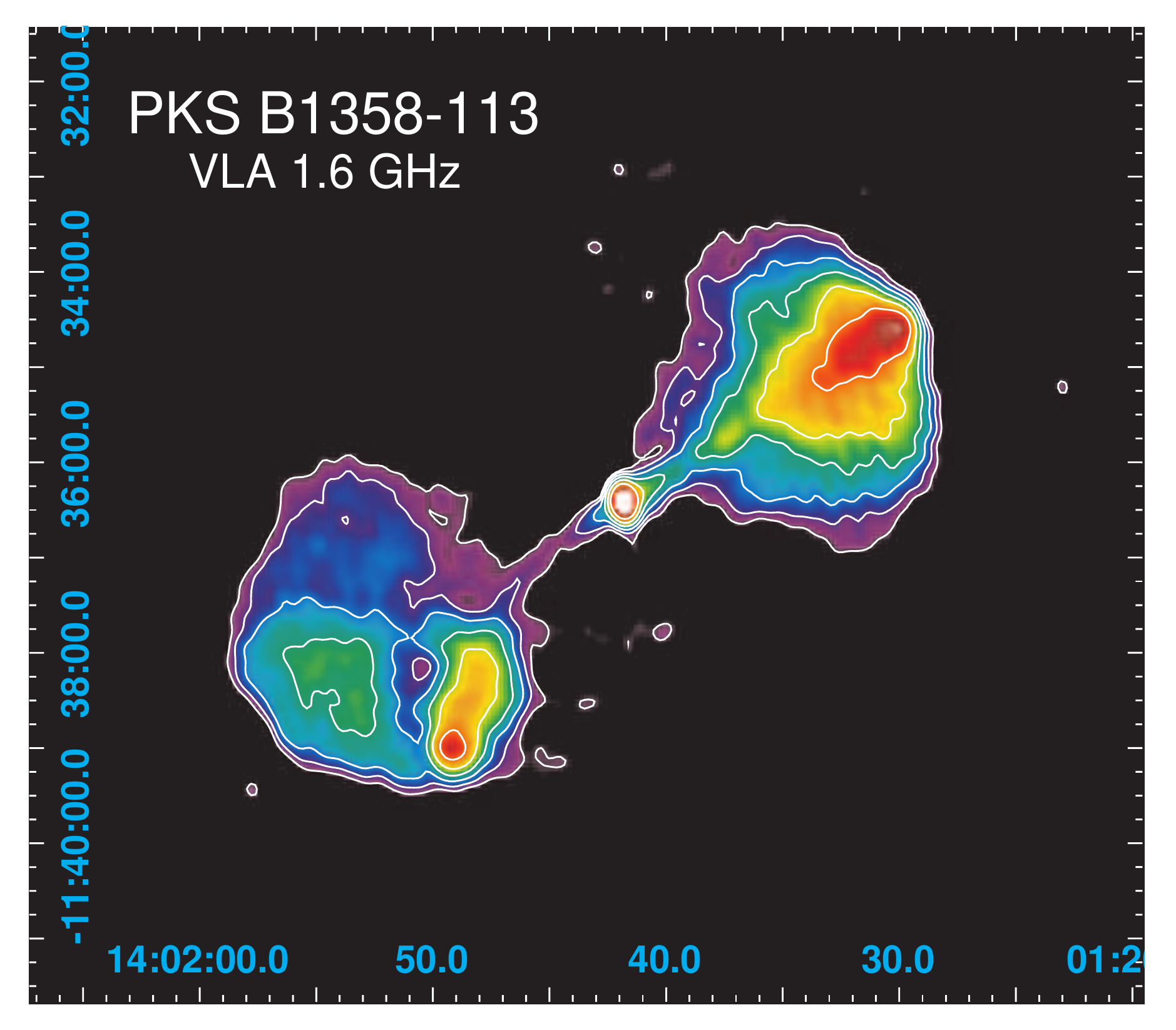}
\caption{\small \vla\ 1.6\,GHz image of \rg\ in the \cl\ cluster in logarithmic scaled colors with R.A. and Decl. (J2000.0 equinox) axes (beam\,=\,15.2\as\,$\times$\,10.9\as, $P\!A = 8.4\deg$). Overlaid are contours beginning at 0.65\,mJy\,beam$^{-1}$ and increasing by factors of two up to 20.6\,mJy\,beam$^{-1}$.}
\label{f-radio}
\end{center}
\end{figure}

We note that prior to the \cxo\ and \xmm\ pointings requested by us, no good-quality X-ray data were available for \rg. The galaxy was located in the \ros\ PSPC field of the cluster Abell\,1837, however offset by 30\am\ from the PSPC aimpoint, so that only the south-east portion of the radio source was visible in the image (the effective PSPC exposure of $11.2$\,ksec), while its north-west portion was obscured by the PSPC support structure. Also, the analysis of the {\it Einstein} Observatory data for \cl\ resulted only in the upper limits for the $0.5-4.5$\,keV cluster luminosity $<1.2 \times 10^{43}$\,erg\,s$^{-1}$ \citep{jon99}. The host galaxy of \rg\ was observed with Hubble Space Telescope (\hst) by \citet{lai03} and \citet{dal09}; we have acquired additional spectroscopic data using the William Herschel Telescope (\wht) at the Roque de los Muchachos Observatory in La Palma (Spain).

\subsection{Archival \vla\ Data}
\label{radio}

To compare with the new X-ray data, we reanalyzed the archival \vla\ data for \rg/\cl\ published originally by \citet{owe92}. These consisted of B-array 1.4\,GHz observations obtained in 1981 June 08 (program: OWEN) and C-array data at 1.6\,GHz from 1983 April 14 (program: AB022). We used AIPS for the standard calibration and DIFMAP \citep{she94} for self-calibration and imaging. The resultant maps have resolution of 5\as\ for the B-array data and 15.2\as\,$\times$\,10.9\as\ (position angle $P\!A = 8.4\deg$) for the C-array. A 4.9\,GHz dataset with 870s exposure from 1988 February 19 (CnB array; program AC208) was also analyzed, but only the radio core was detected so this was not used. The 1.6\,GHz image of \rg\ displaying a characteristic FR\,II morphology is shown in Figure\,\ref{f-radio}.

\begin{figure}[!t]
\begin{center}
\includegraphics[width=\columnwidth]{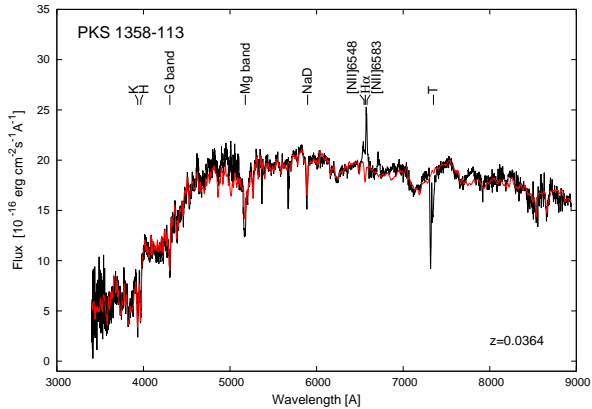}
\caption{\small Rest frame optical spectrum of \rg\ obtained with \wht. Black line denotes the observed spectrum, while the STARLIGHT model fit is plotted in red; telluric lines are marked with `T'.}
\label{f-opt}
\end{center}
\end{figure}

To determine the average radio spectrum of the lobes, we utilized lower resolution maps at 74\,MHz (85\as) from the \vla\ Low-Frequency Sky Survey redux \citep[VLSSr;][]{lan12} and at 1.4\,GHz (45\as) from the NVSS \citep{con98}. At 74\,MHz we measured 5.29\,Jy for the east lobe and 8.10\,Jy for the west lobe; at 1.4\,GHz we measured 0.69\,Jy (east) and 1.18\,Jy (west). Assuming $10\%$ uncertainties in the fluxes, the two band spectral indices for the lobes are indistinguishable with $\alpha_{\rm rad} = 0.69 \pm 0.05$ (east) and $0.66 \pm 0.05$ (west). Utilizing integrated flux measurements at 80 and 160\,MHz from \citet{sle95}, together with 0.41, 1.41, and 2.7\,GHz integrated fluxes from \citet{wri90}, we derive $\alpha_{\rm rad} = 0.67 \pm 0.05$ for the entire source (assuming $15\%$ uncertainties in the individual measurements), consistent with the above values for the individual lobes.

\subsection{New \wht\ Observations}
\label{optical}

The \wht\ observations of \rg\ were carried out in service mode on 2011 April 21. Long-slit spectra were obtained using the Intermediate-dispersion Spectrograph and Imaging System (ISIS), which is the double-armed spectrograph enabling simultaneous observations in both blue and red wavelength ranges. The chosen slit width was 1\as, centered on the nucleus of the source and positioned along the jet axis of the radio galaxy ($P\!A = 139\deg$). In the blue arm the grism R300B was used and the detector was a thinned, blue-sensitive EEV12, consisting of an array of $4096 \times 2048$ (13.5 micron) pixels with a spatial scale of $0.20$\as\ per pixel. The binning factor was $1\times1$ which yielded a wavelength coverage between about $3300\,\AA$ and $5300\,\AA$, with a dispersion of $0.86\,\AA$ per pixel. The instrumental resolution was $3\,\AA$ (FWHM) corresponding to $\sigma_{\rm inst} \simeq 90$\,km\,s$^{-1}$. In the red arm the grism R158R with an order blocking filter GG495 was used. The detector was the default chip for the ISIS red arm (RED+), which is a sensitive array of $4096 \times 2048$ (15.0 micron) pixels with spatial scale 0.224\as\ per pixel. The binning factor was $1\times 1$ which yielded a wavelength coverage between $5300\,\AA$ and $9000\,\AA$, with a dispersion of $1.82\,\AA$ per pixel. The instrumental resolution was $6\,\AA$ corresponding to $\sigma_{\rm inst} \simeq 102$\,km\,s$^{-1}$.

\begin{figure}[!t]
\begin{center}
\vspace{-0.25cm}
\includegraphics[width=\columnwidth]{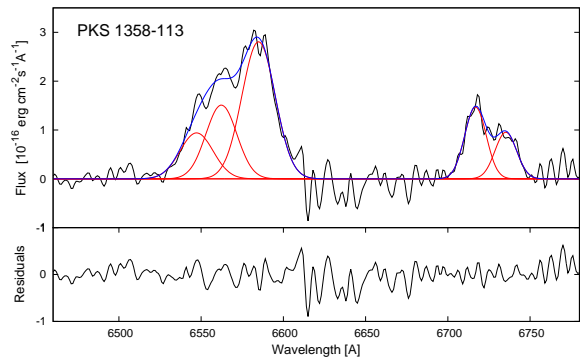}
\caption{\small The \rg\ optical spectrum around the [NII]\,$\lambda 6548, 6584 + {\rm H}\alpha$ blended lines and [SII]\,$\lambda 6717, 6731$ lines (in black). 
Gaussian profiles fitted to the emission lines are plotted in red and the reconstructed blend is given by the blue line while the residuals with respect to the observed data are shown in the bottom panel.}
\label{f-lines}
\end{center}
\end{figure}

The observations were split into two exposures to remove cosmic rays. Spectra of the standard SP1257+038 were obtained for the spectrophotometric calibration. Arc lamp spectra were taken before and after every exposure to allow an accurate wavelength calibration. The average seeing FWHM during the observing run was about 0.8\as. The exposure time was $2 \times 900$\,s per arm, but since the second exposure was made through thin clouds, we analyzed only the first set of data.

Reduction steps were performed for both spectral ranges separately using the NOAO IRAF\footnote{IRAF is distributed by the National Optical Astronomy Observatory, which is operated by the Association of Universities for Research in Astronomy (AURA) under cooperative agreement with the National Science Foundation.} package. A master bias frame was created by averaging all the bias frames obtained during the observing night and subtracted from the science frames. Also a master flat field frame was created and all 2D science frames were corrected for flat field. Then the cosmic rays were removed from the science exposures. Wavelength calibration was performed using ArNe lamp exposures and checked using sky lines; correction for optical distortion was applied. The contribution from the sky was determined from the empty fields and subtracted. Since the slit was positioned along the radio jet axis, which is perpendicular to the photometric major axis, light from a region of 44\as\,$\times$\,1\as\ was summed into a 1D spectrum using the APALL task.

Before fitting of the stellar continuum, 1D spectra were corrected for the Galactic extinction with the maps of \citet{sch98} using the extinction law of \citet{car89}, shifted to the rest frame and resampled to $\Delta \lambda = 1\,\AA$. The analysis of galaxy stellar continuum was performed using the STARLIGHT code \citep{cid05} which allows an observed spectrum to be fit with a linear combination of simple stellar populations (SSPs) extracted from the models of \citet{bru03}. Bad pixels and emission lines were masked and left out of the fits. There were two bases of SSPs used in these fits: (1) $N_{\star}=150$ with 25 ages between 1\,Myr and 18\,Gyr and six metallicities from 0.005 to $2.5\,Z_{\odot}$, and (2) $N_{\star} =45$ with 15 ages between 1\,Myr and 13\,Gyr and three metallicities from 0.004 to $2.5\,Z_{\odot}$. The results of stellar populations and velocity dispersions obtained from the fits performed with these two bases were consistent, implying a mix of 1.43, 11, and 13\,Gyr stellar populations in the system, with metallicities of 0.02 and $0.05\,Z_{\odot}$; we did not find any trace of stellar populations younger than 1\,Gyr. 

\begin{table}[!t]
{\footnotesize
\caption{Emission Lines in the WHT Optical Spectrum of \rg}
\label{t-lines}
\begin{center}
\begin{tabular}{cccc}
\hline\hline
Line & Flux$^{\ast}$ & Position & FWHM\\
	& [10$^{-15}$ \Su]	& [$\AA$]	&	[km\,s$^{-1}$]	\\
\hline
$[$OIII$]$\,$\lambda 4959$ & 1.31 & 4961.2 & 841.0 \\
$[$OIII$]$\,$\lambda 5007$ & 3.90 & 5009.8 & 841.0 \\
$[$NII$]$\,$\lambda 6548$ & 2.37 & 6547.2 & 1084.7 \\
H$\alpha$ & 3.57 & 6562.3 & 1013.4 \\
$[$NII$]$\,$\lambda 6584$ & 7.12 & 6584.9 & 1084.7 \\
$[$SII$]$\,$\lambda 6716$ & 2.41 & 6716.8 & 685.6 \\
$[$SII$]$\,$\lambda 6735$ & 1.56 & 6735.2 & 685.6 \\
\hline\hline
\end{tabular}
\end{center}
$^{\ast}$ Values not corrected for extinction.}
\end{table}

The velocity dispersion for the \rg\ host obtained from our STARLIGHT fits is $\sigma = 295.0 \pm 6.8$ km\,s$^{-1}$. The rest-frame optical spectrum of \rg\ along with the STARLIGHT fit is presented in Figure\,\ref{f-opt}. Emission line fluxes were measured by fitting Gaussian profiles to the residual spectra obtained after subtraction of the stellar light (see Table\,\ref{t-lines} and Figure\,\ref{f-lines}). Fitting was done using the SPECFIT task in the STSDAS external IRAF package. The [NII]\,$\lambda 6548, 6584 + {\rm H}\alpha$, and [SII]\,$\lambda 6717, 6731$ lines were fitted simultaneously. Lines of the same ion were assumed to have the same offset and width, and the additional constraints were further imposed on the flux ratios: [OIII]\,$\lambda 5007$/[OIII]\,$\lambda 4959 = 2.97$ and [NII]\,$\lambda 6584$/[NII]\,$\lambda 6548 = 3$.

\subsection{New \xmm\ Observations}
\label{xmm}

We observed the galaxy cluster \cl\ with \xmm\ on 2010 January 17 (revolution 1851) with a total exposure time of 36\,ks. The pointing was centered on \rg\ hosted by the BCG. The calibrated event files were produced using the \xmm\ Science Analysis System (SAS) version 11. The chip number 4 of EPIC/MOS1 was excluded from the analysis due to the anomalously high level of instrumental background \citep{sno08}. The removal of soft proton flares, the detection and removal of point sources, the modeling of the quiescent particle background, and the extraction of the EPIC data products (images, lightcurves, and spectra) along with instrument response matrices and ancillary files, were all handled by the \xmm\ Extended Source Analysis Software (ESAS) following the methods described in \citet{kun08} and \citet{sno08}\footnote{See also \burl{ftp://legacy.gsfc.nasa.gov/xmm/software/xmm-esas/xmm-esas.pdf}}. Inspection of images after standard ESAS analysis revealed additional point sources undetected by the {\tt cheese} task. With the use of our {\it Chandra} observations (see \S\,\ref{cxo} below) we confirmed the presence of five additional point sources corresponding to the XMM spots; we removed these additional sources and repeated the analysis.

\begin{figure}[!t]
\begin{center}
\includegraphics[width=\columnwidth]{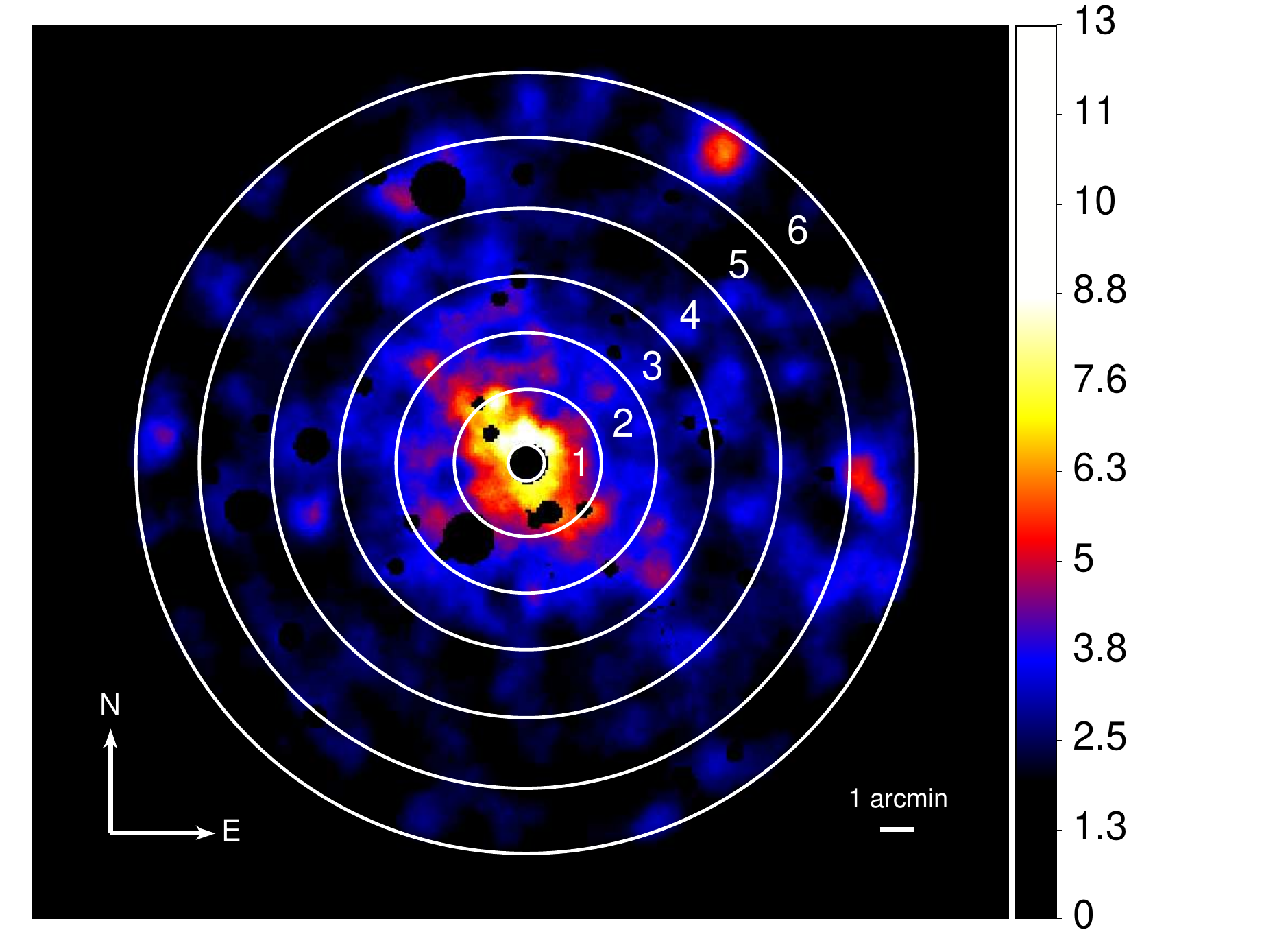}
\caption{\small Combined \xmm\ count image (all three EPIC instruments) of the \cl\ cluster, with the overlaid white contours of six concentric annuli chosen for the analysis of the cluster diffuse emission. The image is CR background-subtracted, exposure-corrected, and adaptively binned. Dark spots correspond to point sources which were subtracted in the prior steps to the analysis.}
\label{f-annuli}
\end{center}
\end{figure}

\subsubsection{Backgrounds and Foregrounds}

The emission of the cluster is contaminated by several components of the EPIC particle and photon foreground and background (FaB). The background spectrum due to interactions of high energy cosmic rays with the telescope has two components: a continuum emission due to particles interacting directly with the detector, and fluorescent emission lines from the material surrounding the detector. The spectrum of the continuum component is modeled in the framework of the ESAS package by tasks {\tt mos\_back} and {\tt pn\_back}, and as such can be directly subtracted from the data before the fitting in the X-ray spectral fitting package {\tt xspec}. The spectrum of the two fluorescent lines changes with the position on the detector and thus needed  to be modeled for each position on the detector and instrument independently.

The particle background due to soft solar protons (SP) interacting with the detector is unpredictable, being composed of fast, high intensity flares, and a quiescent component of longer, low intensity flares. The periods of strong SP flaring were removed from the gathered data based on the inspection of the lightcurves. The cleaning procedure is implemented within the ESAS framework tasks {\tt mos-filter} and {\tt pn-filter}. After this cleaning, about 20\,ks of good time remained for EPIC/MOS1, 20.4\,ks for EPIC/MOS2 and 14\,ks for EPIC/PN. The quiescent component of SP can hardly be distinguished from the real photon events in the source lightcurves. We tested our data for residual contamination by low intensity SP flares using the method of \cite{del04}, where a ratio $r$ of area corrected $8-12$\,keV count rate in the field of view (FOV) to that outside of FOV is calculated. The higher the ratio is, the more significant the residual contamination by SP. In the cases of EPIC/MOS1 and EPIC/MOS2 we obtained $r=1.088\pm0.033$ and $r=1.040\pm0.027$, respectively, which imply that the EPIC/MOS detectors are not contaminated by SP. For EPIC/PN the ratio turned out to be slightly elevated, $r = 1.223\pm0.040$, which indicates some remaining residual contamination by SP. 

The EPIC photon FaB is composed of thermal and nonthermal components. The foreground thermal emission comes from the Local Hot Bubble (LHB) and from the absorbed Galactic Halo (GH). The nonthermal background component is an absorbed composite spectrum of all unresolved extragalactic point X-ray sources (EPL). The model of the EPIC photon FaB needs to be included in the spectral fitting. 

\begin{table}[!t]
{\footnotesize
\caption{Number of background-subtracted counts in the selected annuli and regions for all three \xmm\ instruments} 
\label{t-annuli}
\begin{center}
\begin{tabular}{ccccc}
\hline\hline
Annulus$^{\dagger}$ & Radii  & MOS1 & MOS2 & PN\\
\hline
1 & 32\as$-$1\am10\as & 1085$\pm$39 & 1248$\pm$42 & 1847$\pm$52 \\
2 & 1\am10\as$-$3\am50\as & 1676$\pm$55 & 2009$\pm$61 & 2895$\pm$66 \\
3 & 3\am50\as$-$5\am30\as & 1938$\pm$67 & 2418$\pm$74 & 3203$\pm$77 \\
4 & 5\am30\as$-$7\am30\as & 1621$\pm$61 & 2604$\pm$75 & 3681$\pm$91 \\
5 & 7\am30\as$-$9\am35\as & 1669$\pm$61 & 3111$\pm$81 & 4565$\pm$100 \\
6 & 9\am35\as$-$11\am30\as & 1848$\pm$66 & 3353$\pm$87 & 4670$\pm$106 \\
\hline
\hline
Region$^{\ddagger}$ & Size & MOS1 & MOS2 & PN \\
\hline
A  & 110\as\,$\times$\,90\as & 255$\pm$17 & 234$\pm$17 & 484$\pm$25 \\
B  & 110\as\,$\times$\,90\as & 138$\pm$14 & 201$\pm$16 & 276$\pm$19 \\
C  & 35\as & 434$\pm$22 & 560$\pm$25 & 1010$\pm$33 \\
D  & 75\as & 238$\pm$19 & 271$\pm$20 & 283$\pm$22 \\
E  & 76\as & 280$\pm$21 & 360$\pm$23 & 440$\pm$27 \\
A+B & ---  & 396$\pm$23 & 468$\pm$24 & 796$\pm$32 \\
\hline\hline
\end{tabular}
\end{center}
$^{\dagger}$ See Figures\,\ref{f-annuli} and \ref{f-regions}.\\
$^{\ddagger}$ See Figure\,\ref{f-regions}.}
\end{table}

\subsubsection{Diffuse Emission of the Cluster}

To calculate the temperature, density and pressure profiles of the cluster, we chose six concentric annuli centered on the \rg\ radio galaxy core position, extending out to 11.5\am, as shown in Figure\,\ref{f-annuli}. Table\,\ref{t-annuli} lists the number of background-subtracted counts in each annulus.  The complete model used to fit the XMM data in the annuli, hereafter {\bf Model\,1}, is of the form $C_1 C_2 \, [LHB+ABS\, (GH+EPL+CL)]+ L_1 + L_2$, where $C_1$ is a constant fixed at the value of a solid angle of a region from which the spectrum is extracted, while $C_2$ is a cross calibration constant between the instruments, which is set to unity for MOS1. Galactic absorption ($ABS$) is modeled with {\tt wabs} using Solar abundances of \cite{and89}; thermal components of the foreground (LHB, GH) and cluster emission (CL) per square arcmin in the sky are modeled using {\tt apec}, while the nonthermal component (EPL) is modeled with a power-law function. The temperatures and normalizations of all the thermal components were allowed to vary, but we fixed their abundances at the Solar value. In addition, we fixed the photon index of the EPL component at $\Gamma = 1.41$ \citep{del04}, and the Galactic hydrogen column density in the direction of the radio galaxy, $5 \times 10^{20}$\,cm$^{-2}$ \citep{kal05}. Finally, $L_1$ denotes the fluorescent K line of Al at $\sim 1.5$\,keV, and $L_2$ denotes the fluorescent K line of Si at $\sim 1.75$\,keV; both of these are narrow and were fitted with {\tt gaussian} (note that the latter line is not present in EPIC/PN).

We used {\tt xspec} version 12.6.0 \citep{arn96} to fit three EPIC spectra from each annulus, in the energy band restricted to $0.4-7.0$\,keV. As a background file in {\tt xspec}, we used the spectrum of cosmic-ray continuum calculated for each annulus individually by ESAS (see above). Before fitting, we grouped channels in each spectrum to a minimum of 30 counts per energy bin. We used the $\chi^2$ fit statistic to quantify goodness of the modeling, and fitted the data from each annulus individually. To get the best estimate of the photon FaB, we first modeled the spectrum of the sixth, outermost annulus, where the cluster emission is expected to be the faintest and low with respect to the FaB. For that fit, we fixed the cluster redshift at $z=0.0363$ and abundances at 30\% of the Solar values. The parameters of the best-fit model can be found in Table\,\ref{t-cluster}. Note that the resulting flux of the EPL component in the $2-10$\,keV energy band, $\simeq 2.41 \times 10^{-11}$\,erg\,cm$^{-2}$\,s$^{-1}$\,deg$^{-2}$, is in agreement with the results of \cite{del04} who obtained $\simeq (2.24\pm0.16) \times 10^{-11}$\,erg\,cm$^{-2}$\,s$^{-1}$\,deg$^{-2}$ based on the \ros\ data.

\begin{figure}[!t]
\begin{center}
\includegraphics[width=\columnwidth]{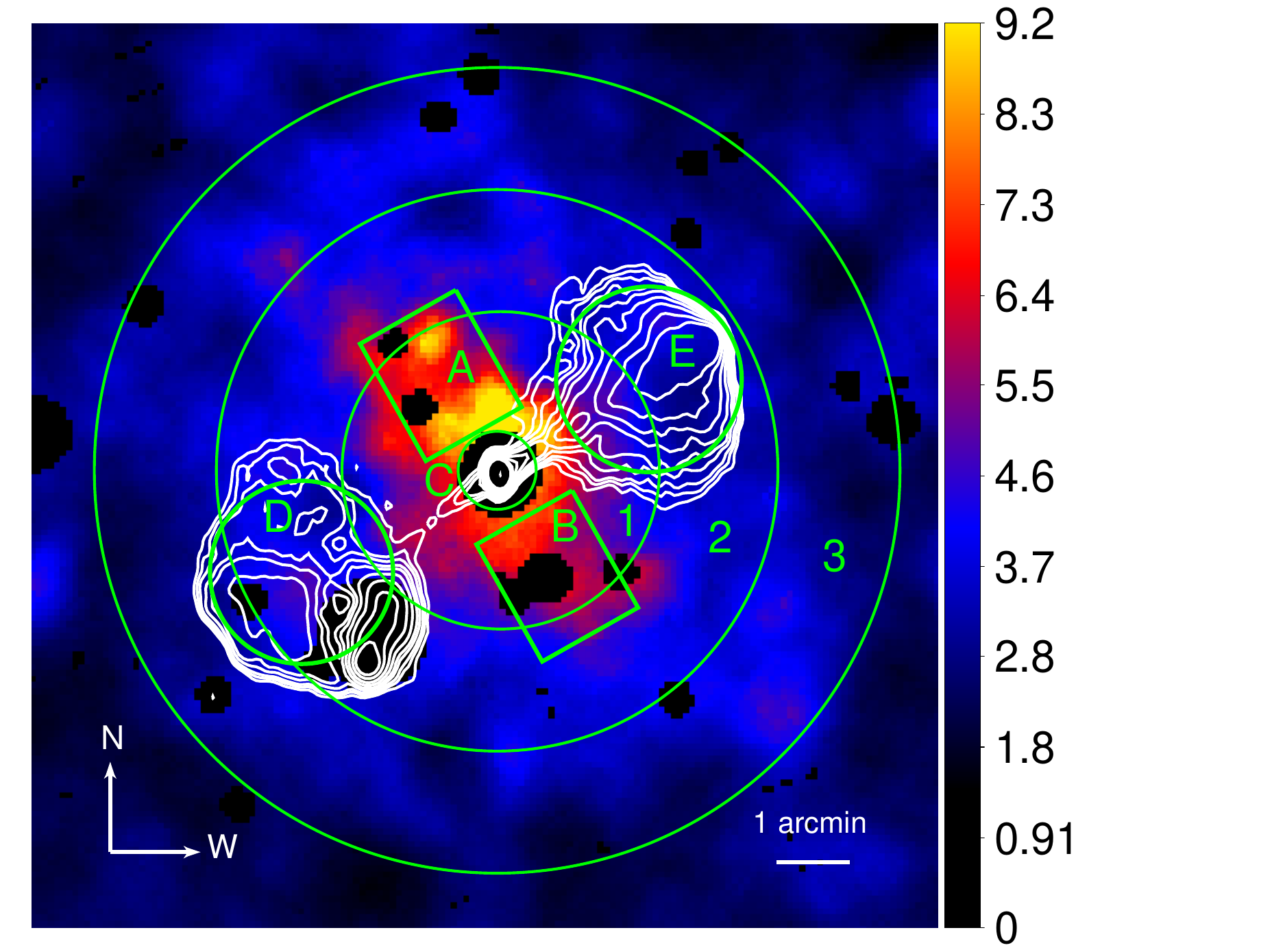}
\caption{\small Combined \xmm\ count image of the central parts of the \cl\ cluster (color scale), with the overlaid 1.6\,GHz emission of \rg\ measured with \vla\ (thick white contours; logarithmic scale). The first three concentric annuli chosen for the analysis of the cluster diffuse emission, as well as the regions A--E selected for the analysis of the X-ray emission around \rg\ radio source, are denoted with thin green contours.}
\label{f-regions}
\end{center}
\end{figure}

In order to fit the spectra of the remaining annuli 1--5, we used the best-fit model of the sixth annulus for fixing all the FaB parameters. We also set the values of $C_1$ to solid angles calculated for each instrument and annulus; $C_2$ for MOS1 was set to 1. The remaining free parameters are the temperature and normalization of the cluster emission, cross calibration constants $C_2$ for MOS2 and PN, and the emission lines which are known to vary with the location on the detector. The obtained best-fit parameters are given in Table \ref{t-cluster}. As follows, the applied Model\,1 provides a good representation of the spectra in annuli 2--6, for which we do not see significant residuals at any energy, indicating that the ICM temperature is increasing in the central parts of the cluster. However, the model is not a good representation of the data in the first annulus. As is evident in Figures\,\ref{f-annuli} and \ref{f-regions}, the cluster emission in this region is not uniform and symmetric, but instead elongated perpendicular to the radio jet axis. The temperature and abundance of this feature, analogous to the one found in, e.g., Cygnus\,A system \citep{smi02,wil06} may be different from those of the surrounding ICM; an additional non-thermal component due to the extended lobes of \rg\ in this region may be an additional issue.

\begin{table*}[!t]
{\footnotesize
\begin{center}
\caption{Best-fit Model\,1$^{\dagger}$ parameters for the diffuse emission of the \cl\ cluster (annuli 1--6)} 
\label{t-cluster}
\begin{tabular}{c c r r r r r r}
\hline\hline
Component & Parameter$^{\ast}$ & 1 & 2 & 3 & 4 & 5 & 6\\
\hline
MOS1 & $C_1$ & 12.22 & 28.29 & 44.05 & 52.03 & 63.58 & 77.31\\
MOS2 & $C_1$ & 12.08 & 28.70 & 45.05 & 69.21 & 98.87 & 120.17\\
PN & $C_1$ & 11.13 & 24.48 & 41.24 & 69.36 & 92.91 & 111.79 \\
MOS1 & $C_2$ & 1 & 1 & 1 & 1 & 1 & 1 \\
MOS2 & $C_2$ & 1.10 & 1.11 & 1.12 & 1.03 & 0.89 & 0.93\\
PN & $C_2$ & 1.12 & 1.22 & 1.11 & 1.01 & 1.07 & 1.16\\

LHB & $kT$ [keV] & 0.01$^{(f)}$ & 0.01$^{(f)}$ & 0.01$^{(f)}$ & 0.01$^{(f)}$ & 0.01$^{(f)}$ & $0.10\pm0.01$ \\
LHB & $A$ [$\times 10^{-6}$] & 3.98$^{(f)}$ & 3.98$^{(f)}$ & 3.98$^{(f)}$ & 3.98$^{(f)}$ & 3.98$^{(f)}$ & $3.98^{+0.72}_{-1.06}$ \\
GH & $kT$ [keV] & 0.26$^{(f)}$ & 0.26$^{(f)}$ & 0.26$^{(f)}$ & 0.26$^{(f)}$ & 0.26$^{(f)}$ & $0.26^{+0.03}_{-0.04}$ \\
GH & $A$ [$\times 10^{-6}$] & 1.07$^{(f)}$ & 1.07$^{(f)}$ & 1.07$^{(f)}$ & 1.07$^{(f)}$ & 1.07$^{(f)}$ & $1.07^{+0.31}_{-0.19}$ \\
EPL & $\Gamma$ & 1.4$^{(f)}$ & 1.4$^{(f)}$ & 1.4$^{(f)}$ & 1.4$^{(f)}$ & 1.4$^{(f)}$ & 1.4$^{(f)}$ \\
EPL & $A$ [$\times 10^{-6}$] & 1.02$^{(f)}$ & 1.02$^{(f)}$ & 1.02$^{(f)}$ & 1.02$^{(f)}$ & 1.02$^{(f)}$ & $1.02^{+0.20}_{-0.09}$ \\

CL & $kT$ [keV] & $1.63^{+0.07}_{-0.08}$ & $1.70^{+0.18}_{-0.10}$ & $1.53^{+0.11}_{-0.17}$ & $1.29^{+0.17}_{-0.25}$ & $1.09^{+0.22}_{-0.04}$ & $1.29^{+0.71}_{-0.21}$ \\
CL & $Z/Z_{\odot}$ & 0.3$^{(f)}$ & 0.3$^{(f)}$ & 0.3$^{(f)}$ & 0.3$^{(f)}$ & 0.3$^{(f)}$ & 0.3$^{(f)}$\\
CL & $A$ [$\times 10^{-6}$] & $17.4_{-2.2}^{+2.0}$ & $8.62_{-1.33}^{+0.99}$ & $4.79_{-0.98}^{+0.72}$ & $1.42_{-0.58}^{+0.31}$ & $0.77_{-0.28}^{+0.26}$ & $1.15_{-0.39}^{+0.23}$ \\
\hline
 & $\chi^2$ & 183.71 & 207.18 & 279.72 & 372.71 & 463.75 & 579.79\\
 & dof & 142 & 242 & 317 & 385 & 465 & 538\\
 & $\chi^2_{\nu}$ & 1.293 & 0.856 & 0.882 & 0.96 & 0.997 & 1.078\\
\hline\hline
\end{tabular}
\end{center}
$^{\dagger}$ Model\,1 is of the form $C_1 C_2 \, [LHB+ABS\, (GH+EPL+CL)]+ L_1 + L_2$.\\
$^{\ast}$ The normalizations of the $C_1$ and $C_2$ constants are provided per arcmin$^2$.}
\end{table*}

\subsubsection{Extended Vicinities of \rg }

Because significant departures from spherical symmetry may produce spurious oscillations in the deprojected temperature and density profiles discussed below in the following \S\,\ref{s-dep}, first we performed detailed modeling of selected spatial regions in the extended vicinity of \rg. We selected five new spectral regions, including the northern and southern parts of the aforementioned elongated X-ray feature (regions A and B), east and west lobes (regions D and E), and the innermost parts of the cluster including the central AGN (region C). These are shown in Figure\,\ref{f-regions} (see also Table\,\ref{t-annuli}). Regions A and B did not contain sufficient counts to be fitted independently, therefore we only fitted the combined region A+B. The spectra in regions A+B, D and E were fitted with three different models: {\bf Model\,1} specified in the previous paragraph, {\bf Model\,2} in which the thermal (cluster) emission was replaced with a power law component, $C_1 C_2 \, [LHB+ABS\,(GH+EPL+PL)]+ L_1 + L_2$, and {\bf Model\,3} which combines the thermal and power-law components, $C_1 C_2 \, [LHB+ABS\,(GH+EPL+PL+CL)]+ L_1 + L_2$. The results of these fits are summarized in Table\,\ref{t-fits}.

\begin{figure}[!t]
\begin{center}
\includegraphics[width=\columnwidth]{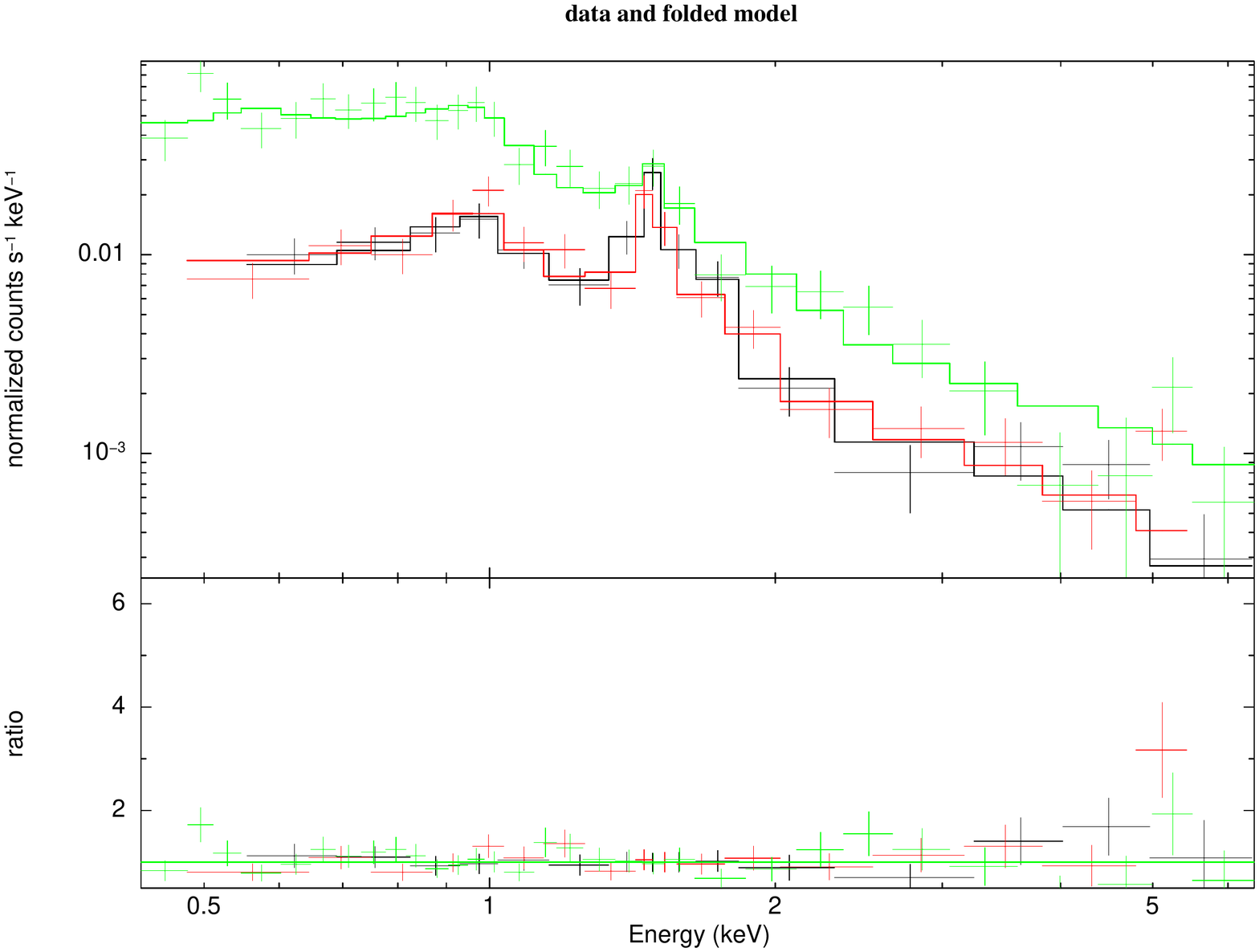}
\caption{\small \xmm\ spectrum of the region A+B (MOS1 -- black, MOS2 -- red, PN -- green), along with the Model\,3 fit (see \S\,\ref{xmm} for the description).}
\label{f-AB}
\end{center}
\end{figure}

In the case of the Model\,1 applied to the spectrum of region E there is a large uncertainty in the temperature determination despite a good quality of the fit. The Model\,2 with only a power-law component provides a very good description of the data, with the photon index well constrained. The Model\,3 is also acceptable in the statistical sense, although the model parameters are only poorly constrained. The size of the region D is similar to that of region E, but the number of counts in region D is lower due to the presence of a bright point source excluded from the analysis, as well as the presence of PN chip breaks and bad pixel columns. Hence we were not able to obtain reasonable fits to the region D spectrum with any of our models. The data from the region A+B are best represented with Model\,3. Here the thermal component --- which as we believe corresponds to the gaseous filaments embedded within the oldest parts of the radio lobes, in the aforementioned analogy to Cygnus\,A --- seems needed to model the data below 2\,keV, while the PL component --- which is most likely due to the inverse-Compton emission of the section of the lobes overlapping with the selected regions --- is required by the data above 2\,keV. The \xmm\ spectrum of region A+B along with the Model\,3 fit is shown in Figure\,\ref{f-AB}.

\begin{figure}[!t]
\begin{center}
\includegraphics[width=\columnwidth]{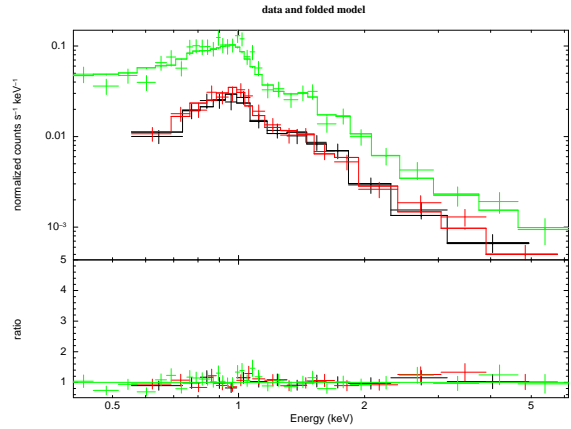}
\caption{\small \xmm\ spectrum of the cluster central region C (MOS1 -- black, MOS2 -- red, PN -- green), along with the Model\,4 fit (see \S\,\ref{xmm} for the description).}
\label{f-core}
\end{center}
\end{figure}

The spectrum of the innermost parts of the \cl\ cluster, including the \rg\ nucleus, was extracted from the central region of 35\as\ radius. This relatively large extraction region C was chosen to assure that the number of counts is sufficient to handle the FaB. As a background file we used the cosmic ray background continuum calculated the same way as the background for all the other analyzed regions. First, we modeled the data with our basic Model\,1. This model provides unacceptable fits to the data with  $\chi^2_{\nu}>2$ and large positive residuals at energies above 2\,keV and around 0.8\,keV. The residuals around 0.8\,keV suggest a multi-phase structure of the inner parts of the cluster, while the residuals above 2\,keV indicate an additional non-thermal component. We therefore used, rather for illustrative purposes only, a cooling flow (CF) model {\tt cflow} of \citet{mus88}, in which the thermal emission arises due to a multi-temperature gas cooling radiatively. This model provides a very good fit to the data below 2 keV. In order to deal with the residuals above 2\,keV, we add to this model also a power law component, and fit the entire spectrum of region C with {\bf Model\,4} of the form $C_1 C_2 \, [LHB+ABS\,(PL+GH+EPL+CF)+L_1+L_2$. In the fitting we were forced to fix the cluster abundance, since otherwise the fits did not converge. Here we chose $Z = 0.6 \, Z_{\odot}$ noting that the central regions of clusters have typically higher metallicity than the outer regions. The results of the fit are summarized in Table \ref{t-fits}, and shown in Figure\,\ref{f-core}. The fact that the lower temperature returned by the CF fit is relatively high, $\sim 0.7$\,keV, suggests that the applied simple cooling flow model does not provide any realistic representation of the innermost parts of the \cl\ cluster \citep[see the related discussion in][]{wer13}. The limited photon statistics preclude us however from any further more sophisticated modeling.
 
\begin{table*}[!t]
{\scriptsize
\begin{center}
\caption{Best-fit Model\,1--4$^{\dagger}$ parameters for the X-ray emission associated with \rg\ (regions E, A+B, and C)} 
\label{t-fits}
\begin{tabular}{ccccccc}
\hline\hline
Region & Component & Parameter$^{\ast}$ & Model\,1 & Model\,2 & Model\,3 & Model\,4 \\
\hline
E (west lobe) & MOS1 & $C_1$ & 4.92 & 4.92 & 4.92 & --\\
 & MOS2 & $C_1$ & 5.02  & 5.02 & 5.02 & --\\
& PN & $C_1$ & 4.02  & 4.02 & 4.02 & --\\
&PL & $\Gamma$ & -- & $2.05^{+0.33}_{-0.29}$ & $2.0^{+5.9}_{-4.0}$  & --\\
&PL & $A$ [$\times 10^{-6}$] & -- & $2.58^{+0.82}_{-0.85}$ & $2.0^{+1.2}_{-2.0}$ & --\\
&PL & $F_{\rm 0.3-10\,keV}$\,[$ \times 10^{-14}$] & -- & $1.41_{-0.44}^{+0.45}$ & $1.15_{-1.06}^{+0.60}$ & --\\
&CL & $kT$ [keV] & $2.48^{+2.13}_{-0.87}$ & -- & $1.3^{+1.3}_{-1.3}$ & --\\
&CL & $Z/Z_{\odot}$ & 0.3$^{(f)}$ & -- & 0.3$^{(f)}$ & --\\
&CL & $A$ [$\times 10^{-6}$] & $9.5^{+2.8}_{-3.6}$ & -- & $1.57^{+9.6}_{-1.6}$ & --\\
&CL & $F_{\rm 0.3-10\,keV}$\,[$ \times 10^{-14}$] & $1.0_{-0.41}^{+0.45}$ & -- & $0.14_{-0.14}^{+1.37}$ & --\\
& & $\chi^2$ & 33.49 & 31.52 & 30.95 & -- \\
& & dof & 34 & 34 & 32 & -- \\
& & $\chi^2_{\nu}$ & 0.98 & 0.93 & 0.97 & --\\
\hline
A+B & MOS1 & $C_1$ & 4.28 & 4.28 & 4.28 & --\\
 & MOS2 & $C_1$ & 4.05 & 4.05 & 4.05  & --\\
 & PN & $C_1$ & 4.01 & 4.01 & 4.01 & --\\
 & PL & $\Gamma$ & -- & $2.07^{+0.15}_{-0.22}$ & $1.52^{+0.39}_{-1.08}$ & -- \\
 & PL & $A$ [$\times 10^{-6}$] & -- & $6.2^{+0.7}_{-1.6}$ & $1.90^{+1.3}_{-1.5}$  & --\\
 & PL & $F_{\rm 0.3-10\,keV}$\,[$ \times 10^{-14}$] & -- & $3.34_{-0.66}^{+0.38}$ & $1.56_{-0.61}^{+0.55}$  & --\\
 & CL & $kT$ [keV] & $1.30^{+0.37}_{-0.08}$ & -- & $1.17^{+0.13}_{-0.15}$ & -- \\
 & CL & $Z/Z_{\odot}$ & 0.3$^{(f)}$ & -- & 0.3$^{(f)}$  & --\\
 & CL & $A$ [$\times 10^{-6}$] & $13.0^{+6.5}_{-2.7}$ & -- & $8.8^{+4.1}_{-3.0}$  & --\\
 & CL & $F_{\rm 0.3-10\,keV}$\,[$ \times 10^{-14}$] & $1.30_{-0.27}^{+0.28}$ & -- & $0.93_{-0.29}^{+0.38}$  & --\\
&  & $\chi^2$ & 68.46 & 91.24 & 47.35  & --\\
& & dof & 47 & 47 & 45  & --\\
 & & $\chi^2_{\nu}$ & 1.46 & 1.94 & 1.05 & -- \\
\hline
C (center) & MOS1 & $C_1$ & -- & -- & -- &1.00 \\
& MOS2 & $C_1$ & -- & -- & -- &1.02 \\
& PN & $C_1$ & -- & -- & -- &1.07 \\
& CFLOW & $kT_{\rm low}$ [keV] & -- & -- & -- &$0.68^{+0.09}_{-0.07}$ \\
& CFLOW & $kT_{\rm high}$ [keV] & -- & -- & -- &$1.51^{+0.29}_{-0.26}$ \\
& CFLOW & $Z/Z_{\odot}$ & -- & -- & -- &0.6$^{(f)}$ \\
& CFLOW & $A$ [M$_{\odot}$ yr$^{-1}$] & -- & -- & -- & $2.6^{+1.4}_{-0.6}$ \\
& CFLOW &  $F_{\rm 0.3-10\,keV}$\,[$ \times 10^{-13}$] & -- & -- & -- &$1.16_{-0.11}^{+0.21}$ \\
& PL & $\Gamma$ & -- & -- & -- &$0.94^{+0.44}_{-0.62}$ \\
& PL & $A$ [$\times 10^{-6}$] & -- & -- & -- &$5.9^{+5.6}_{-3.7}$ \\
& PL &  $F_{\rm 0.3-10\,keV}$\,[$ \times 10^{-13}$] & -- & -- & -- & $1.10_{-0.25}^{+0.25}$ \\
& & $\chi^2$ & -- & -- & -- &37.74 \\
& & dof & -- & -- & -- &54 \\
& & $\chi^2_{\nu}$ & -- & -- & -- &0.70 \\
\hline\hline
\end{tabular}
\end{center}
$^{\dagger}$ Model\,1 is of the form $C_1 C_2 \, [LHB+ABS\, (GH+EPL+CL)]+ L_1 + L_2$; Model\,2 is $C_1 C_2 \, [LHB+ABS\,(GH+EPL+PL)]+ L_1 + L_2$; Model\,3 is $C_1 C_2 \, [LHB+ABS\,(GH+EPL+PL+CL)]+ L_1 + L_2$; Model\,4 is $C_1 C_2 \, [LHB+ABS\,(PL+GH+EPL+CF)+L_1+L_2$.\\
$^{\ast}$ The normalizations of the $C_1$ constant are provided per arcmin$^2$. Fluxes $F_{\rm 0.3-10\,keV}$ are given in the units of \Su\,arcmin$^{-2}$.}
\end{table*}

\subsection{New \cxo\ Observations}
\label{cxo}

We observed the \cl\ cluster with the \cxo\ X-ray Observatory on 2010 May 26 (obsid=11750). The central radio galaxy \rg\ was placed on the Advanced CCD Imaging Spectrometer (ACIS) and was offset from the default aimpoint position in order to locate the cluster central regions on the ACIS-I CCD array (chips ACIS-01236 were active). The observation was performed in VFAINT mode which allows for efficient background event filtering. The standard deadtime correction was applied during the processing resulting in the total exposure time of 58.817\,ks. We used CIAO version 4.4 software \citep{fru06} to perform the \cxo\ data analysis. We processed the data using {\tt chandra\_repro} and applied the calibration files from the CALDB version 4.4 including the ACIS model contamination file {\tt acisD1999-08-13contamN0006.fits}. The sub-pixel algorithm EDSER was also applied giving us the best spatial resolution X-ray image. All spectral extraction was performed with {\tt specextract} and modeling was done in {\it Sherpa} \citep{fre01}. All spectral models were fit to the data in the $0.5-7$\,keV energy range.

\subsubsection{Cluster Center and \rg\ Nucleus}

The central regions of the \cl\ cluster are well resolved in the \cxo\ image. We used CHART\footnote{\burl{http://cxc.harvard.edu/chart/}} to simulate the PSF centered on the central galaxy assuming the X-ray spectrum and the normalization based on the X-ray galaxy nucleus model (see below). The PSF simulations indicate that a 2.5\as\ circular region contains 96.5\% of the photons from a central point source. Therefore, in the analysis of the cluster emission we used only the regions outside of the inner 2.5\as\ radius, which were not contaminated by the BCG, up to 30\as\ from the center; this region is embedded within the \xmm\ center region C (see \S\,\ref{xmm} above). We fit the spectrum assuming an absorbed {\tt apec} plasma model with the Galactic absorbing column density of $5 \times 10^{20}$\,cm$^{-2}$ as before. In this way we obtained the best-fit temperature of $kT = 2.0^{+0.1}_{-0.1}$\,keV and the metal abundances of $Z = 0.55^{+0.22}_{-0.16} \, Z_{\odot}$.

We also extracted the X-ray spectrum of the \rg\ nucleus assuming a circular region with 1.5\as\ radius, as appropriate for a point source emission in \cxo\ observations. This region contains a total of $312 \pm18$ counts. The typical background contamination of a point source observed with \cxo\ is negligible, however, in our case the AGN is embedded in the X-ray cluster emission which contributes to the total extracted spectrum. We therefore assumed the background region as the 2.5\as--30\as\ inner annulus discussed above. We extracted the background spectrum and calculated the background contribution using the appropriate scaling between the source and background region sizes; after applying the background correction we were left with $309.5 \pm17.8$ net counts within the $0.5-7$\,keV energy range. We next assumed an absorbed power law model for the AGN X-ray emission, setting the absorption column density at the Galactic value. The fit returned the power law photon index of $\Gamma=2.1\pm0.1$, and the unabsorbed fluxes of $F_{\rm 0.5-2.0\,keV} = (2.90 \pm0.25) \times10^{-14}$\Su\ and $F_{\rm 2-10\,keV} = (3.1 \pm0.6) \times10^{-14}$\Su.

\subsection{X-ray Deprojection}
\label{s-dep}

Based on the \xmm\ data analysis described in section \ref{xmm} above, we used the {\tt xspec} model {\tt projct} to perform a cluster de-projection into spherical shells of constant (by assumption) electron density defined by the set of previously selected annuli 1--6 (see Table \ref{t-annuli} and Figure\,\ref{f-annuli}). This allows us to estimate cluster parameters in 3-D space from the 2-D projected \xmm\ spectra, assuming the absorbed {\tt apec} model for the ICM emission\footnote{\burl{http://heasarc.nasa.gov/xanadu/xspec/manual/XSmodelProjct.html}}. Since the MOS1 is missing two of its chips, we perform the de-projection using only the MOS2 and PN data. The results of the de-projection are summarized in Table \ref{t-dep}, where the uncertainties are provided at the 90\% level for one significant parameter.

In order to determine the temperature profile in the innermost parts of the \cl\ cluster, we also extracted the \cxo\ spectra from three sub-annuli $i-iii$ with inner--outer radii of 2.5\as--5\as, 5\as--15\as, and 15\as--30\as, respectively. We set the background annulus as 60\as--80\as. Next we fit the spectra of the three sub-annuli in {\it Sherpa} assuming an absorbed {\tt apec} plasma model and using the {\tt deproject} task\footnote{\burl{http://cxc.harvard.edu/contrib/deproject/}} \citep[see][]{sie10}. Both the Galactic absorption column density and the metallicity were kept constant during the fit. The resulting temperatures, normalizations, and electron densities are listed in Table~\ref{t-dep}, where the uncertainties are provided at the 90\% level for one significant parameter. Figure\,\ref{f-deplog} includes the extension of the temperature, density, and pressure profiles of the \cl\ cluster measured with \xmm\ to the innermost cluster regions probed with \cxo.

\section{Results of the Analysis}
\label{results}

\subsection{\rg\ Radio Galaxy}
\label{rg}

Despite its clear FR\,II-like large-scale radio morphology, \rg\ is not particularly bright at radio frequencies. The total 1.4\,GHz radio flux density of its extended structure, 1.87\,Jy (see \S\,\ref{radio}), implies a spectral power of $P_{\rm 1.4\,GHz} \simeq 0.6 \times 10^{25}$\,W\,Hz$^{-1}$, which is below the Fanaroff-Riley dividing line. Additionally, the host of \rg\ is a particularly bright galaxy, with a $R$-band absolute magnitude of $M_{\rm R} \simeq -23.0$ \citep[see][converted using the 158\,Mpc distance to the source adopted here]{owe89,lai03}. For this host galaxy magnitude, the Ledlow-Owen division corresponds to the critical value of $P_{\rm 1.4\,GHz} \sim 3 \times 10^{25}$\,W\,Hz$^{-1}$ \citep[see][]{wol07,che09}, hence \rg\ appears \emph{under-luminous in radio} considering its large-scale radio morphology and starlight luminosity.

\begin{table*}[!t]
{\scriptsize
\begin{center}
\caption{De-projected model parameters for the diffuse X-ray emission$^{\dagger}$ of the inner (sub-annuli $i-iii$) and outer (annuli 1--6) parts of the \cl\ cluster based on the \cxo\ and \xmm\ data analysis, respectively} 
\label{t-dep}
\begin{tabular}{c ccc ccc ccc}
\hline\hline
Parameters & $i$ & $ii$ & $iii$ & 1 & 2 & 3 &4 & 5 & 6 \\
\hline
$kT$ [keV] & $1.08^{+0.07}_{-0.04}$ & $1.67^{+0.17}_{-0.11}$ & $1.72^{+0.24}_{-0.11}$ & $1.29^{+0.95}_{-0.38}$ & $2.72^{+1.26}_{-0.67}$ & $1.81^{+0.45}_{-0.22}$ & $1.0^{+0.49}_{-0.28}$ & $1.08^{+0.26}_{-0.27}$ & $1.60^{+0.95}_{-0.38}$ \\
$A$ [$\times 10^{-4}$] &  $77.4^{+7.4}_{-6.4}$ & $ 3.57^{+0.34}_{-0.30}$ & $0.42^{+0.04}_{-0.05}$ & $1.0{\pm 0.17}$ & $2.18{\pm0.33}$ & $3.12\pm0.49$ & $0.92^{+0.73}_{-0.39}$ & $1.02\pm0.61$ & $5.06^{+0.97}_{-0.96}$ \\
$n_e$ [$10^{-4}$\,cm$^{-3}]$ &  $461^{+22}_{-19}$ & $ 99^{+5}_{-4}$ & $ 34^{+2}_{-2}$ & $5.02^{+0.40}_{-0.44}$ & $3.34^{+0.25}_{-0.27}$ & $2.66^{+0.20}_{-0.22}$ & $0.94^{+0.32}_{-0.23}$ & $0.74^{+0.20}_{-0.27}$ & $1.40^{+0.13}_{-0.13}$ \\
$p_{\rm th}$ [$10^{-12}$\,dyn\,cm$^{-2}$] & $79.8^{+6.4}_{-6.4}$ &   $26.5^{+3.0}_{-3.0}$ & $9.4^{+1.4}_{-1.4}$ & $2.0^{+1.7}_{-0.7}$ & $2.8^{+1.6}_{-0.9}$ & $1.47^{+0.50}_{-0.28}$ & $0.29^{+0.29}_{-0.13}$ & $0.24^{+0.14}_{-0.13}$ & $0.68^{+0.51}_{-0.21}$ \\
\hline\hline
\end{tabular}
\end{center}
$^{\dagger}$ Assuming the absorbed {\tt apec} model.}
\end{table*}

Based on detailed \hst\ Space Telescope Imaging Spectrograph and ground-based spectroscopy, augmented by \hst\ imaging, \citet{dal09} derived the mass of the SMBH in \rg\ as $M_{\rm BH} = 3.61^{+0.41}_{-0.50} \times 10^9 \, M_{\odot}$. They noted that this rather extreme mass --- \emph{in fact one of the largest measured dynamically} ---  is inconsistent (at the $3\sigma$ level) with the values $\sim 10^9 \, M_{\odot}$ predicted by the $M_{\rm BH} - L_{\rm bulge}$ and $M_{\rm BH} - \sigma$ relations \citep{fer05}, and $\sim 0.6 \times 10^9 \, M_{\odot}$ implied by the `BH fundamental plane' \citep{hop07}. \citet{mcc11} revised the black hole mass determination in the system, refining it to $M_{\rm BH} = 3.9^{+0.4}_{-0.6} \times 10^9 \, M_{\odot}$, which we adopt hereafter. The velocity dispersion measured in our \wht\ data, $\sigma \simeq 295.0 \pm 6.8$ km\,s$^{-1}$ (see \S\,\ref{optical}), is in agreement with the previous \hst-based determination of $\sigma \simeq 288 \pm 14$\,km\,s$^{-1}$ \citep{dal09,mcc11}.

Our best fit to the \wht\ data reveal only the narrow component of the H$\alpha$ line \citep[cf.,][]{dal09}. We find $\log[{\rm NII}]/{\rm H}\alpha \simeq 0.3$, $\log[{\rm OIII}]/{\rm H}\alpha \simeq 0.04$, and ${\rm EW_{H\alpha}} \simeq 1.8\,\AA$. These values, according to \citet{kew06} and \citet{cid10}, indicate that the \rg\ nucleus is a LINER type (`low-ionization nuclear emission-line region') AGN. The measured H$\alpha$ luminosity of $L_{\rm H\alpha} \simeq 10^{40}$\,\Lu\ translates to an AGN bolometric luminosity of $L_{\rm nuc} \simeq 2 \times 10^3 L_{\rm H\alpha} \simeq 2 \times 10^{43}$\,\Lu\ \citep[see][]{sik13}, signaling a \emph{very low accretion rate} in the system, $\Lambda \equiv L_{\rm nuc}/L_{\rm Edd} \simeq 4 \times 10^{-5}$, for the corresponding Eddington luminosity $L_{\rm Edd} \simeq 5 \times 10^{47}$\,\Lu. This is consistent with the X-ray ($0.5-10$\,keV) luminosity of the \rg\ core measured with \cxo\ (see \S\,\ref{cxo}), namely $L_{\rm nuc,\,X} \simeq 2 \times 10^{41}$\,\Lu\,$\simeq 4 \times 10^{-7} \, L_{\rm Edd}$ \citep[in this context, see][]{rus13}.

The mass accretion rate in the nucleus of \rg\ can be estimated as
\begin{equation}
\dot{m} \equiv {\dot{M}_{\rm acc} \over \dot{M}_{\rm Edd}} = \Lambda \, \left({\eta_{\rm d} \over 0.1}\right)^{-1} \, ,
\end{equation}
where $\eta_{\rm d}$ is the radiative efficiency of the accretion disk. According to \citet{sch07}, at low accretion rates in the range $10^{-4} \lesssim \dot{m} \lesssim 10^{-2}$, the $\eta_{\rm d}$ parameter is roughly constant at the few percent level. Applying this value to the case of \rg, we obtain $\dot{M}_{\rm acc} \sim 2 \times 10^{-4} \, \dot{M}_{\rm Edd} \sim 0.02 \, M_{\odot}$\,yr$^{-1}$, the accretion luminosity $L_{\rm acc} \equiv \dot{M}_{\rm acc} c^2 \sim 10^{45}$\,\Lu\,$\simeq 2 \times 10^{-3}\,L_{\rm Edd}$, and the maximum expected jet kinetic power \citep[see][]{mck12} roughly as $L_{\rm j\,(max)} \simeq 3 \, \dot{M}_{\rm acc} c^2 \sim 3 \times 10^{45}$\,\Lu .

The 1.4\,GHz monochromatic luminosity of the extended emission in \rg, $L_{\rm 1.4\,GHz} \simeq 8 \times 10^{40}$\,\Lu, together with the derived radio spectral index $\alpha_{\rm rad} \simeq 0.67$ (see \S\,\ref{radio}), gives the total radio luminosity at the level of $L_{\rm rad} \sim 10^{42}$\,\Lu. Let us therefore comment here on the scaling between the jet kinetic power and the source radio luminosity discussed by \citet{bir08}, \citet{cav10}, and \citet{osu11} in the context of radio lobes in clusters of galaxies. Using the relations derived in these papers, one may find the expected (for a given $L_{\rm 1.4\,GHz}$) jet kinetic power in \rg\ as $L_{\rm j\,(rad)} \sim (1-3) \times 10^{44}$\,\Lu. This is about one order of magnitude lower than $L_{\rm j\,(max)}$ estimated above. In the analysis of the aforementioned works however, the discussed scalings were derived based on the `cavity powers', i.e., the enthalpy of the lobes $\mathcal{H}_{\ell} = 4 \, p_{\ell} V_{\ell}$ --- evaluated for a given lobes volume $V_{\ell}$ and pressure $p_{\ell}$ equal by assumption to the thermal pressure of the surrounding ICM, $p_{\rm th}$ --- divided by the lobes' lifetime approximated by the sound-crossing (or buoyancy) timescale $\tau_{s}$, namely $L_{\rm j\,(cav)} \sim 4 \, p_{\rm th} V_{\ell}/\tau_s$.

In the case of \rg, the total linear size of the radio structure is $L\!S \simeq 450$\as\,$\simeq 320$\,kpc. The total lobes' volume in the system then reads as $V_{\ell} = \pi \, L\!S^3 / 4 \, A\!R^2 \sim 3 \times 10^{71}$\,cm$^3$ for the lobes' axial ratio $A\!R \simeq 1.5$ as observed in the \vla\ maps (see Figure\,\ref{f-radio}) and in agreement with what is expected for an FR\,II-type source. The thermal pressure of the ICM around the radio lobes of \rg\ (annulus 3; see \S\,\ref{xmm}) is roughly $p_{\rm th} \sim 10^{-12}$\,dyn\,cm$^{-2}$. Hence $4 \, p_{\rm th} V_{\ell} \sim 10^{60}$\,erg. Furthermore, for the temperature of the \cl\ ICM around $kT \sim 1$\,keV, that is for the sound velocity of about $c_s = (5 \, kT / 3 \mu \, m_p)^{1/2} \sim 5 \times 10^7$\,cm\,s$^{-1}$, the sound-crossing timescale reads as $\tau_s \sim L\!S/ 2 \, c_s \sim 300$\,Myr. Therefore, $L_{\rm j\,(cav)} \sim 10^{44}$\,\Lu, which is indeed close to the `radio-scaled' value $L_{\rm j\,(rad)}$.

The sound crossing timescale is however not a reliable proxy of the jet lifetime in the specific system analyzed here because, as we argue below, the expansion of over-pressured lobes in \rg\ is supersonic, meaning $p_{\ell} > p_{\rm th}$ and the jet lifetime $\tau_{\rm j} < \tau_s$. For the same reason, the total energy delivered by the jets over $\tau_{\rm j}$ is expected to be larger than the enthalpy $\mathcal{H}_{\ell}$ (which is the sum of the lobes' internal energy and the $p\, dV$ work done on the surrounding medium) because some fraction of this energy should additionally be deposited in a bow shock driven in the surrounding medium by the expanding lobes. As we discuss in the next section, the collected X-ray data provide evidence for the presence of such a shock. Note also that in the case of a source lifetime exceeding 100\,Myr, one should expect the lobes' continuum spectra at GHz frequencies  to be very steep due to radiative cooling of the radio-emitting electrons. However, the measured radio slope of \rg\ is relatively flat up to 3\,GHz, implying the spectral age $\tau_{\rm j} < 100$\,Myr for the expected --- as argued below --- lobes' magnetic field intensity $B_{\ell} > 1$\,$\mu$G \citep[see, e.g.,][]{kon06,mac07}.

The analysis of the \xmm\ data from regions E and A+B coinciding with the \rg\ cavity suggests the presence of a non-thermal emission component with photon index $\sim 2$ and  $0.5-10$\,keV flux of $\sim 10^{-14}$\,\Su\ (see \S\,\ref{xmm} and Table\,\ref{t-fits}). Taking into account the relatively low photon statistics and complex structure of the central parts of the \cl\ cluster, we do not consider this a robust detection, but instead a conservative upper limit for non-thermal X-ray emission related to the lobe \citep[see in this context][]{har10}. Assuming no leakage of low-energy cosmic ray electrons from the lobes \citep[see][]{pfr13}, the corresponding monochromatic luminosity $L_{\rm 1\,keV} \leq 10^{40}$\,\Lu\ can therefore be compared with the expected 1\,keV luminosity of the western lobe resulting from inverse-Compton up-scattering of CMB photons by the radio-emitting electrons,

\begin{equation}
L_{\rm 1\,keV} \sim L_{\rm 1.4\,GHz} \, {u_{\rm CMB} \over u_B} \, \left({B_{\ell} \over 300\,{\rm \mu G}}\right)^{1-\alpha_{\rm rad}} \,
\end{equation}
where $u_{\rm cmb} \simeq 4 \times 10^{-13}$\,\Uu\ is the energy density of the CMB radiation and $u_B \equiv B_{\ell}^2/ 8 \pi$ is the magnetic field energy density. This comparison implies $B_{\ell} \geq 3$\,$\mu$G. Moreover, we note that assuming $B_{\ell} \sim 3 $\,$\mu$G and the ratio of the energy densities stored in the lobes' magnetic field and ultrarelativistic electrons $u_{e\pm}/u_B \sim 10$, as expected for a FR\,II-type system \citep{kat05,cro05,iso11}, the resulting bolometric synchrotron radio luminosity of the system turns out to be of the order of the observed radio power $L_{\rm rad} \sim 10^{42}$\,\Lu. 

For these reasons, we consider $B_{\ell} \sim 3$\,$\mu$G as a rather realistic estimate, leading to $\tau_{\rm j} < 80$\,Myr.  Hence, the jet kinetic power $L_{\rm j} > 4 \, p_{\rm th} V_{\ell} / \tau_{\rm j} > 5 \times 10^{44}$\,\Lu\,$\sim 10^{-3} L_{\rm Edd}$,  and appears larger than the value expected from scaling relations utilizing radio fluxes and cavity powers \citep{bir08,cav10,osu11}, and indicating \emph{the jet production efficiency in \rg\ is close to the maximum expected level} $L_{\rm j\,(max)}$ for the given accretion rate of $\dot{M}_{\rm acc} \sim 0.02 \, M_{\odot}$\,yr$^{-1}$. 

In this context, one can ask if the accretion rate derived here based on the AGN emission line properties is of the order of the Bondi accretion rate, or instead is much lower/higher. According to our analysis of the \cxo\ data (\S\,\ref{cxo}), the gas number density increases between $r \sim 20$\,kpc and $r \sim 2$\,kpc from the \cl\ core roughly as $n_e(r) \propto r^{-3/2}$ (see Table\,\ref{t-dep}). Still, the innermost region resolved with \cxo\ is about one order of magnitude larger than the accretion radius in the system, $r_A = 2 \, G M_{\rm BH} \, c_s^{-2} \sim 200$\,pc (assuming the inner gas temperature of $kT \sim 0.7$\,keV; see Table\,\ref{t-fits}). Hence, one may speculate that either {\bf (i)} the gas density remains constant within the central region at the $n_e \sim 0.04$\,cm$^{-3}$ level measured with \cxo\ around $\sim 2$\,kpc, or {\bf (ii)} the gas number density continues to increase as $\propto r^{-3/2}$ down to the accretion radius scale, reaching in particular $n_e(r_{\rm A}) \simeq 2$\,cm$^{-3}$. In the former case, the Bondi accretion rate,
\begin{equation}
\dot{M}_{\rm Bon} \simeq 4 \pi \, m_p n_e \, G^2 M_{\rm BH}^2 \, c_s^{-3} \, ,
\end{equation}
gives $\sim 0.03 \, M_{\odot}$\,yr$^{-1}$, which should be considered as a strict lower limit for this parameter.\footnote{We note that with $\dot{M}_{\rm Bon} \sim 0.03 \, M_{\odot}$\,yr$^{-1}$, the scaling relations of \citet{all06} and \citet{mer07} between the Bondi power $L_{\rm Bon} = 0.1 \, \dot{M}_{\rm Bon} c^2$ and the jet kinetic luminosity return $L_{\rm j\,(Bon)} \sim 3 \times 10^{43}$\,\Lu, which is much below our lower limit for $L_{\rm j}$.} In the latter case, the Bondi rate turns out about two orders of magnitude higher, $\sim 2.5 \, M_{\odot}$\,yr$^{-1}$, being almost exactly the same as the mass deposition rate returned by fitting the cooling flow model to the \xmm\ data extracted from the innermost parts ($r<20$\,kpc) of the analyzed cluster (see Table\,\ref{t-fits}). However, as cautioned before, the applied cooling flow model does not provide any realistic representation of the multi-phase structure of the gaseous environment within the \cl\ center, and hence this latter value should be considered as a strict upper limit. All in all, the comparison between $\dot{M}_{\rm acc} \sim 0.02 \, M_{\odot}$\,yr$^{-1}$ derived based on the AGN emission line properties, and the crudely estimated $0.03 \, M_{\odot}$\,yr$^{-1}$\,$\ll \dot{M}_{\rm Bon} \ll 2.5 \, M_{\odot}$\,yr$^{-1}$, indicates therefore that \emph{the actual mass accretion rate in \rg\ is below the Bondi value}. This conclusion, on the other hand, is consistent with the expected mass-loss in radiatively inefficient accretion flows which, for radii $r < r_A$, can be approximated by a power-law scaling $ \dot{M}_{\rm acc}\!(r) = \left(r / r_A\right)^{\kappa} \times \dot{M}_{\rm Bon}$ with $\kappa = 0-1$ \citep[see the discussion in][]{kuo14,nem14}.

\subsection{Shock Heating in \cl\ Cluster}
\label{sh}

\begin{figure}[!t]
\begin{center}
\includegraphics[width=\columnwidth]{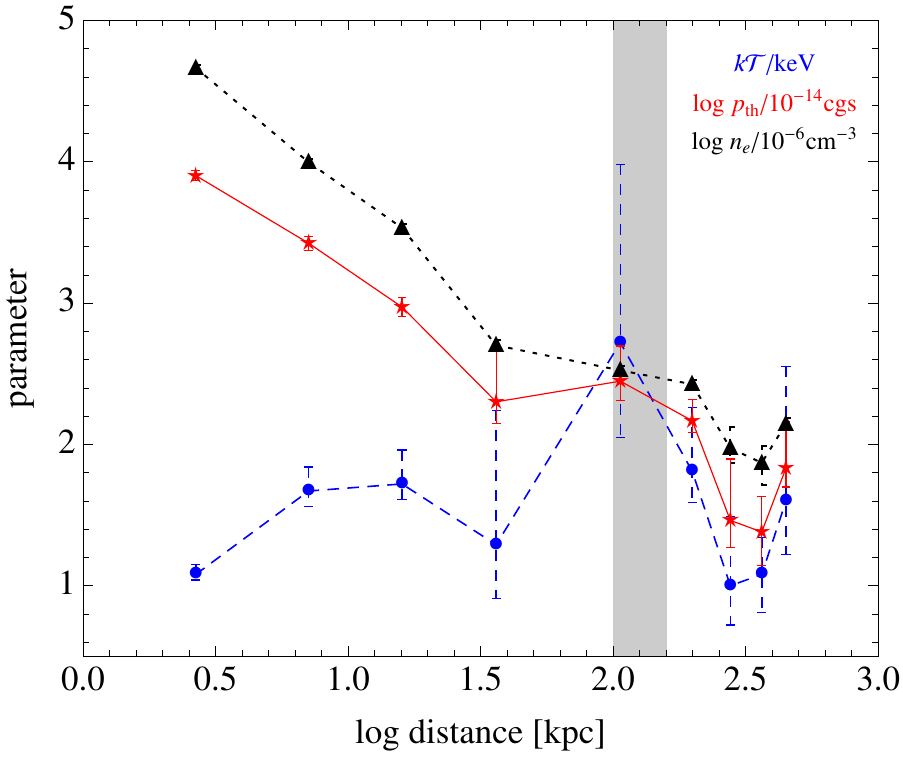}
\caption{\small \cl\ temperature (blue circles), density (black triangles), and pressure (red stars) profiles measured with \xmm\ and \cxo\ (Table~\ref{t-dep}). The position of the edges of the \rg\ radio structure ($\simeq 100$\,kpc in the northeast---southwest direction, $\simeq 160$\,kpc in the northwest---southeast direction) are marked with the gray rectangle.}
\label{f-deplog}
\end{center}
\end{figure}

Our \xmm\ observations of \cl\ reveal a total X-ray luminosity $L_{\rm 0.3-10\,keV} \simeq 4 \times 10^{42}$\,\Lu\ and average gas temperature $kT \simeq 1$\,keV in agreement with the luminosity-temperature correlation derived for clusters and groups of galaxies \citep[e.g.,][]{sun12}. We also found enhancements in the hot gas density and temperature around the outermost edges of the \rg\ radio structure ($\sim 100-160$\,kpc from the cluster center; see Figure\,\ref{f-deplog}). These enhancements are the expected signatures of bow shocks driven in the ICM by the expanding lobes as discussed below.

\begin{figure}[!t]
\begin{center}
\includegraphics[width=\columnwidth]{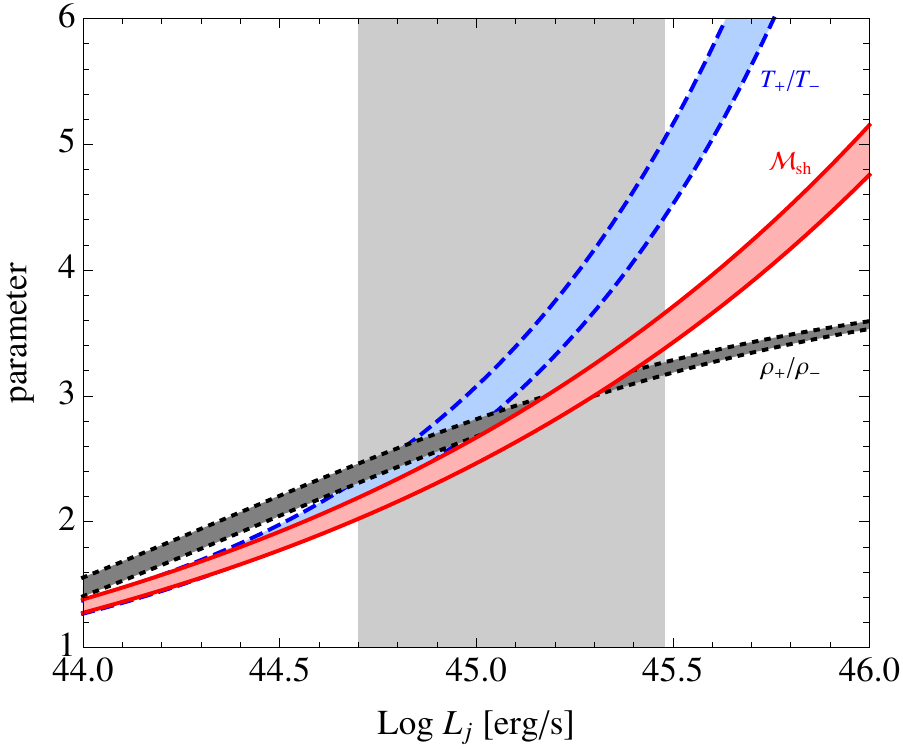}
\caption{\small The model values of the Mach number of a bow shock $\mathcal{M}_{sh}$ around the \rg\ lobes (red solid lines), along with the expected downstream-to-upstream gas density ratio $\rho_+/\rho_-$ (black dotted lines) and temperature ratio $T_+/T_-$ (blue dashed lines), as functions of the jet kinetic power $L_{\rm j}$. The preferred range $L_{\rm j} \simeq (0.5-3) \times 10^{45}$\,\Lu\ is marked with the gray rectangle; the filled areas between the curves correspond to the allowed range $\tau_{\rm j} \simeq 40-70$\,Myr.}
\label{f-sh}
\end{center}
\end{figure}

In the following, we adopt the well-established model for FR\,II radio galaxies evolving in a stratified cluster (or galaxy group) atmosphere with a power-law density profile $\rho(r) = \rho_0 \, (r/r_0)^{-\beta}$, following \citet{kai00} based on \citet{kai97}. This predicts a source linear size
\begin{equation}
L\!S \simeq 190 \, \left({L_{\rm j} \over 10^{45}\,{\rm erg/s}}\right)^{2/7} \left({\tau_{\rm j} \over 30\,{\rm Myr}}\right)^{6/7} {\rm kpc} \, ,
\end{equation}
for the given ICM density profile parameters ($\beta$, $\rho_0$ and $r_0$), the lobes' axial ratio ($A\!R$), and the adiabatic indices for the ICM ($\hat{\gamma}_g$) and jet cocoon ($\hat{\gamma}_{\ell}$). In the above we took $\beta = 3/2$, $\rho_0 = 6.7 \times 10^{-26}$\,g\,cm$^{-3}$ and $r_0 = 2.7$\,kpc (see Table\,\ref{t-dep}), $A\!R = 3/2$ (see \S\,\ref{rg}), $\hat{\gamma}_g = 5/3$ and $\hat{\gamma}_{\ell} = 4/3$. The preferred range for the jet kinetic luminosity $L_{\rm j} \simeq (0.5-3) \times 10^{45}$\,\Lu\ returns the observed value of $L\!S \simeq 320$\,kpc for the source age in a narrow range $\tau_{\rm j} \simeq 40-70$\,Myr, consistent with the upper limit $\tau_{\rm j} < 80$\,Myr derived from the combined radio and X-ray data for \rg\ (see the discussion in \S\,\ref{rg}). For the thus constrained ranges of $L_{\rm j}$ and $\tau_{\rm j}$, the expected model-dependent pressure within the jet cocoon,
\begin{equation}
p_{\ell} \simeq 6 \times 10^{-12} \left({L_{\rm j} \over 10^{45}\,{\rm erg/s}}\right)^{1/7} \left({\tau_{\rm j} \over 30\,{\rm Myr}}\right)^{-11/7} {\rm dyn\,cm^{-2}} ,
\end{equation}
gives $\simeq (2-5) \times 10^{-12}$\,dyn\,cm$^{-2}$, while the Mach number of the bow shock driven in the ICM,
\begin{equation}
\mathcal{M}_{sh} = c_s^{-1} \, \sqrt{p_{\ell} / \rho} \, ,
\end{equation}
turns out to be $\simeq 2-4$ for the 1\,keV-temperature gas. This estimate of $p_{\ell}$ is consistent with the non-thermal pressure in the lobes due to relativistic electrons and magnetic field $(u_{e\pm} + u_B)/3 \gtrsim 10^{-12}$\,dyn\,cm$^{-2}$ corresponding to the anticipated conditions $u_{e\pm} \sim 10 \, u_B$ and $B_{\ell} \gtrsim 3$\,$\mu$G, with sufficient uncertainties to allow for (though not strictly requiring) a substantial contribution of relativistic protons and thermal matter to the total pressure of the jet cocoon \citep[in this context, see][]{dun05,dey06,bir08,osu13,sta13}. Also, the derived Mach number range is consistent with values implied by the density and temperature enhancements (factor of $\sim 2-3$) revealed by the de-projected cluster profiles around the lobes' outermost edges (see Table\,\ref{t-dep} and Figure\,\ref{f-deplog}), though the corresponding emissivity enhancements are not obviously present on the X-ray count images of the \cl\ cluster. This is illustrated in Figure\,\ref{f-sh}, where we plot the expected model values of $\mathcal{M}_{sh}$ along with the corresponding downstream-to-upstream gas density and temperature ratios that follow from the standard Rankine-Hugoniot shock jump conditions as functions of the jet kinetic power.

Weak shocks (Mach numbers $\Ms \lesssim 2$) generated by the supersonic expansion of radio lobes were previously found in the central parts of famous clusters Perseus \citep{fab03,fab06}, MS\,0735.6+7421 \citep{mcn05,git07}, Hercules \citep{nul05her}, Hydra \citep{nul05hyd,wis07,sim09,git11}, Virgo \citep{for05,for07,sim07}, and Cygnus \citep{wil06}. More recently, additional evidence for AGN-driven weak shocks was also reported for  Abell\,4059 \citep{rey08}, RBS\,797 \citep{cav11}, Abell\,2052 \citep{bla11}, and possibly in a poor cluster surrounding 3C\,288 \citep{lal10}.

Similarly weak or somewhat stronger shocks (Mach numbers up to $\Ms \lesssim 4$) were found around radio galaxies residing in poorer (group) environments, including low-power systems NGC 3801 \citep{cro07}, B2\,0838+32A \citep{jet08}, HCG\,62 \citep{git10}, NGC\,5813 \citep{ran11}, and 4C+29.30 \citep{sie12}, as well as the FR\,II source 3C\,452 \citep{she11}. The exceptionally strong shock ($\Ms \sim 8$) has been found around the inner lobe of the Centaurus\,A radio galaxy \citep{kra03,kra07,cro09}. Finally, for completeness we note the $\Ms \lesssim 5$ shocks which have been claimed in the X-ray data gathered for kpc-scale lobes inflated by low-power jets in the halos of Seyfert galaxies Mrk\,6 and Circinus \citep{min11,min12}, and the tentative evidence for an analogous (though much weaker) feature in the Galactic Halo proposed to be related to the Fermi Bubbles \citep{kat13}.

The shocks driven in the ambient medium by expanding AGN jets therefore appear common, and their impact on the surrounding gaseous environments well recognized, particularly in galaxy groups and isolated systems. Still, the observed gas entropy profiles within cluster cores require rather powerful heating sources, which can be reconciled with repeated (on timescales of $\gtrsim 100$\,Myr) shocks driven by jets with relatively large kinetic powers $L_j \gtrsim 10^{45}$\Lu\ that additionally last for relatively long periods of at least $\tau_j \sim 10$\,Myr \citep[see the discussion in, e.g.,][]{voi05,zan05,mat11}. Such are the typical characteristics of high-power FR\,II radio galaxies \citep[see, e.g.,][]{mac07,bird08,ant12}.  Conversely, most radio bubbles found in the centers of rich low-$z$ clusters were formed in rather short periods of low-power FR\,I-type jet activity \citep[see][]{bir04,raf06,dun08}. 

A notable exception to this rule is Cygnus\,A with its archetypal `classical double' morphology, for which the jet lifetime of order $\sim 30$\,Myr and total jet kinetic power $L_j \sim 10^{46}$\Lu\ have been estimated based on various datasets and numerical simulations \citep{car91,sta07,mat10}, in agreement with what is implied by the presence of a $\Ms \sim 1.3$ shock at the boundaries of its $140$\,kpc-scale lobes \citep{wil06}. Another exception is 3C\,444 in the cluster Abell\,3847, with a $\Ms \sim 1.7$ shock surrounding its $\sim 200$\,kpc-scale FR\,II-like radio structure, with $L_j$ and $t_j$ parameters analogous to those derived for Cygnus\,A \citep{cro11}. Only four other systems are comparable to these two examples, but none of them display a clear FR\,II morphology. Hercules\,A, for example, is a transitional FR\,I/FR\,II source with a $\Ms \sim 1.6$ shock found about $\sim 160$\,kpc from the cluster center, suggesting $t_j \gtrsim 30$\,Myr and a relatively high jet power of $L_j \gtrsim 10^{46}$\Lu\ \citep{nul05her}. The other FR\,I/FR\,II system 3C\,288 is less spectacular, with its $\sim 70$\,kpc lobes and $\Ms \sim 1.4$ shock \emph{possibly} related to the lobes' expansion \citep{lal10}. The radio structure of Hydra\,A is complex, consisting of an inner ($<100$\,kpc) and outer ($\sim 300$\,kpc) pair of lobes, both of the FR\,I type \citep{tay90,lan04,wis07}; the $\Ms \sim 1.3$ shock surrounding the outer structure indicates jets with $L_j \sim 3 \times 10^{45}$\Lu\ active for $\gtrsim 100$\,Myr \citep{nul05hyd,sim09}. Finally, the total energy stored in the $\sim 200$\,kpc-scale $\Ms \sim 1.4$ shock front associated with the amorphous radio halo in MS0735+7421 implies jets with $L_j \gtrsim 10^{46}$\Lu\ active for $\sim 100$\,Myr \citep{mcn05,git07}.

The \rg/\cl\ system discussed in this paper may constitute an important addendum to the above list, with its $\mathcal{M}_{sh} \simeq 2-4$ shock driven at $\gtrsim 100$\,kpc distances from the cluster center by the expanding FR\,II-type lobes inflated for about $\tau_{\rm j} \simeq 40-70$\,Myrs by jets with kinetic luminosity $L_{\rm j} \simeq (0.5-3) \times 10^{45}$\,\Lu. The bulk of the total energy transported by these jets, $E_{\rm tot} \simeq 2\,L_{\rm j} \, \tau_{\rm j} \sim (2-8) \times 10^{60}$\,ergs, is deposited into the shock-heated surrounding thermal gas and the internal energy of the jet cavity; only a negligible fraction of $E_{\rm tot}$ is radiated away via non-thermal processes at radio and X-ray frequencies.

The Mach number of the bow shock claimed above is relatively high, keeping in mind the other shocks found so far in clusters of galaxies, although one should emphasize here that \cl\ is a relatively poor system with low gas density and temperature. Let us therefore comment in this context on the fact that while the presence of plasma discontinuities in poorer environments were inferred through analysis of temperature and surface brightness profiles of the X-ray emitting plasma, the corresponding shock Mach numbers were estimated in several cases based solely on the derived temperature enhancements, under the working assumption that these correspond precisely to the temperature jumps following from the macroscopic Rankine-Hugoniot jump conditions for an ideal gas. Yet the `observed' temperature jumps follow from the analysis of the thermal electron emission (bremsstrahlung) continua, and as such, reflect the electron temperatures rather than temperatures of the dynamically dominating ions. This distinction may be of importance, since the electron heating at non-relativistic collisionless shocks is still an open problem, as it is controlled by hardly understood microscopic processes involving a variety of plasma instabilities operating at and around shock fronts. These various processes may proceed differently in different types of shocks, depending on the upstream plasma parameters. Interestingly, \emph{in situ} observations of low- and intermediate-$\mathcal{M}_{sh}$ collisionless shocks in the Solar System indicate that the fraction of the dissipated energy that goes into thermal electron heating decreases with increasing shock Mach number \citep{tho87,sch88,mas11}. 

Let us therefore speculate that the thermal electron heating at bow shocks driven by the expanding AGN jets and lobes in clusters and groups of galaxies is similarly inefficient, with an analogous inverse dependence on the shock Mach number. If true, this would then create \emph{an additional bias against detecting stronger shocks in poorer environments}. The gas density here comes into play because even in the case of inefficient electron heating at the shock, the electron-ion temperature equilibration due to Coulomb collisions is inevitable in the downstream. The related timescale for this process is
\begin{equation}
\tau_{ei} \simeq 0.01 \, {m_p \, (kT_e)^{3/2} \over e^4 \, n_e \, \sqrt{m_e}} \, ,
\end{equation}
which, for the $kT_e \sim 1$\,keV temperature electron gas with number density $n_e \sim 10^{-4}$\,cm$^{-3}$ (i.e., for the parameters of the ICM in \cl\ upstream of the shock), reads as $\tau_{ei} \gtrsim 100$\,Myr. This timescale is longer than the lifetime of the radio source \rg\ ($\tau_{\rm j} < 100$\,Myr), which may imply that the electron and ion temperatures in the shocked ICM did not yet equilibrate fully. On the other hand, in cases of environments denser than that of \cl\ (e.g., M\,87) or in radio galaxies more evolved than \rg\ (e.g., Hydra\,A), the temperature equilibration due to electron-ion Coulomb collisions in the downstream may proceed faster, or at least sufficiently fast, to assure $\tau_{ei} < \tau_{\rm j}$, so that the discussed effect is not important.

If, however, the parameters of a system are such that $\tau_{ei} > \tau_{\rm j}$ (which we expect to happen in the cases of younger radio galaxies evolving in poorer environments), \emph{and} the collective plasma processes responsible for the electron heating at the shock front are indeed inefficient, as hypothesized here, then relatively strong shocks driven by the expanding lobes may appear `weaker' on the temperature maps when compared with the density maps \citep[see in this context the case of the FR\,II radio galaxy 3C\,444 evolving in the Abell\,3847;][]{cro11}. One should emphasize here that the detection of weak shocks in lower-density environments is in general difficult, as the resulting surface brightness enhancement are expected to be rather minor. In addition, the broad band response of the available X-ray detectors is more sensitive to the gas density rather than temperature, which makes temperature jumps rather unreliable proxies for shock Mach numbers (especially taking into account in addition the effects of projection of curved shock fronts onto the sky). Still, the analysis of the temperature maps may in many cases be the only method available for this purpose, in particular in the cases of stronger shocks for which the density jump is not very sensitive to $\mathcal{M}_{sh}$ and saturates quickly at the asymptotic value of $4$ (see Figure\,\ref{f-sh}). The other complication in this respect may be however an increased efficiency of the Fermi-type acceleration of suprathermal electrons with the increasing shock Mach number \citep[see in this context the recent discussion in][]{mas13,kan14,guo14}, resulting in the fact that the X-ray emission from behind high-$\mathcal{M}_{sh}$ cluster shocks may be predominantly non-thermal \citep[as in the case of Cen\,A inner lobes;][]{cro09}.

\section{Summary and Conclusions}
\label{discuss}

In this paper we analyze in detail the gathered radio (\vla), optical (\wht), and X-ray (\xmm\ and \cxo) data for the radio galaxy \rg\ located in the center of the \cl\ cluster. We find that \rg\ appears under-luminous ($L_{\rm 1.4\,GHz} < 10^{41}$\,\Lu) considering its large-scale FR\,II morphology, as well as the starlight luminosity of its host, thus challenging the strictness of the Fanaroff-Riley and Ledlow-Owen divisions in classifying extragalactic radio sources. The host of \rg\ is the brightest cluster galaxy in \cl, harboring one of the most massive black holes known to date, $M_{\rm BH} \simeq 4 \times 10^9 \, M_{\odot}$. Our analysis of the obtained \wht\ data reveals that this black hole is only weakly active, with the corresponding LINER-type nuclear luminosity of $L_{\rm nuc} \sim 4 \times 10^{-5} \, L_{\rm Edd} \sim 2 \times 10^{43}$\,\Lu. Following the results of numerical simulations by \citet{sch07}, we assume an efficient electron heating in the accretion disk in the system; this leads to a relatively high ($\sim 1\%$) radiative efficiency of the disk despite its low accretion rate, estimated here as $\dot{M}_{\rm acc} \sim 2 \times 10^{-4} \, \dot{M}_{\rm Edd} \sim 0.02 \, M_{\odot}$\,yr$^{-1}$ with a corresponding accretion luminosity $L_{\rm acc} \sim 2 \times 10^{-3}\,L_{\rm Edd} \sim 10^{45}$\,\Lu. The derived $\dot{M}_{\rm acc}$ is likely lower than the Bondi value, which we could only restrict to a relatively wide range ($0.03-2.5 \, M_{\odot}$\,yr$^{-1}$\,$\ll \dot{M}_{\rm Bon} \ll 2.5 \, M_{\odot}$\,yr$^{-1}$) because the accretion radius is not resolved in our \cxo\ observation.

Based on the collected radio and X-ray data for \rg, along with the well-established model for the evolution of FR\,II radio galaxies, we derive the preferred range for the jet kinetic luminosity $L_{\rm j} \simeq (1-6) \times 10^{-3}\, L_{\rm Edd} \sim (0.5-3) \times 10^{45}$\,\Lu. This range is significantly above those implied by various scaling relations proposed for radio sources in galaxy clusters, being instead very close to the maximum jet power allowed for the given accretion rate ($\sim 3 \, \dot{M}_{\rm acc} c^2$) thus implying a very high jet production efficiency in a low-accretion rate AGN. This is in agreement with recent results of numerical simulations of jets launched by spinning SMBHs \citep{mck12}. We also constrain the source lifetime as $\tau_{\rm j} \simeq 40-70$\,Myrs, meaning the total amount of deposited jet energy is $E_{\rm tot} \simeq 2\,L_{\rm j} / \tau_{\rm j} \sim (2-8) \times 10^{60}$\,ergs. We argue that about half of this energy goes into shock-heating of the surrounding thermal gas, while about $50\%$ is deposited into internal energy of the jet cavity (while a negligible fraction is radiated away), in agreement with  expectations from recent long-term simulations of relativistic jet evolution \citep{per11}.

The detailed analysis of the obtained X-ray data for the \rg/\cl\ system reveals some evidence for the presence of a bow-shock driven by the expanding radio lobes into the cluster environment. We derive the shock Mach number in the range $\mathcal{M}_{sh} \simeq 2-4$, which is one of the highest claimed for clusters or groups of galaxies. This, together with the recent growing evidence that powerful FR\,II radio galaxies may not be uncommon in the centers of clusters at higher redshifts, strongly supports the idea that jet-induced shock heating may indeed play an important and possibly dominant role in shaping the properties of clusters, galaxy groups, and massive ellipticals in formation. Such a role may be underestimated in studies focused on low-power radio galaxies typically found in the centers of low-redshift systems. At the same time, we note that this `mechanical' feedback in more distant sources may be difficult to access observationally due to several various reasons, including limited sensitivities and resolutions of the available X-ray detectors, but also the fact that the collective plasma processes responsible for electron heating at the shock front may become increasingly inefficient for stronger shocks, as in fact observed in the Solar System. This latter effect, when combined with a relatively long electron-ion Coulomb equilibration timescale, may result in stronger shocks driven by the expanding jets in a lower-density plasma appearing `weaker'.

\section*{Acknowledgments}

{\L}.~S. and M.~O. were supported by Polish NSC grant DEC-2012/04/A/ST9/00083.
A.~Sz. and G.~M. were supported by Chandra grant GO0-11144X.
Work by C.~C.~C. at NRL is supported in part by NASA DPR S-15633-Y. 
Support for A.~S. was provided by NASA contract NAS8-03060.
The authors thank the anonymous referee for her/his critical reading of the submitted manuscript and constructive comments which helped to improve the paper.

{}

\end{document}